\pdfoutput=1

\documentclass[preprint,11pt]{elsarticle}
\usepackage{dsfont}
\usepackage[T1]{fontenc}
\usepackage[english]{babel}
\usepackage{csquotes}
\usepackage{fancyvrb}
\usepackage{amsmath,amssymb}
\usepackage{mathtools}
\usepackage{tensor}
\usepackage{scalefnt}
\usepackage[centertableaux, smalltableaux]{ytableau}
\usepackage{graphicx}
\usepackage{listings}
\usepackage[tableposition=top]{caption}
\usepackage{booktabs}
\usepackage[final]{showkeys} 
\usepackage{acronym}
\usepackage{nth}
\usepackage{imakeidx}
\usepackage{array}
\usepackage{ifthen}
\usepackage{enumitem}
\usepackage[hidelinks,bookmarksnumbered=true,plainpages=false,linktocpage=true]{hyperref}
\usepackage[capitalise]{cleveref}
\biboptions{sort&compress}

\crefname{relation}{relation}{relations}
\creflabelformat{relation}{{\upshape(#2#1#3)}}
\crefformat{appendix}{#2#1#3}
\definecolor{codegreen}{rgb}{.58,.69,.5}
\definecolor{codegray}{rgb}{.5,.5,.5}
\definecolor{codeblue}{rgb}{.35,.39,.6}
\definecolor{codered}{rgb}{.55,.3,.45}
\definecolor{backcolour}{rgb}{.95,.95,.95}
\newcommand\CODEstyle{\color{black}\ttfamily\scriptsize}
\makeatletter
\lstdefinestyle{mystyle}{
  basicstyle=%
    \ttfamily
    \lst@ifdisplaystyle\scriptsize\fi
}
\makeatother
\lstdefinestyle{pythoncode}{
  frame=single,
  language=Python,
  backgroundcolor=\color{backcolour},
  commentstyle=\color{codegreen},
  keywordstyle=\color{codeblue},
  numberstyle=\color{codegray}\ttfamily\tiny,
  stringstyle=\color{codeblue},
  basicstyle=\CODEstyle,
  breakatwhitespace=false,
  breaklines=true,
  postbreak=\mbox{\textcolor{codegray}{$\hookrightarrow$}\space},
  captionpos=b,
  keepspaces=true,
  numbers=left,
    numbersep=5pt,
    showspaces=false,
    showstringspaces=false,
    showtabs=false,
    tabsize=2,
    comment=[l]{\#},
}
\lstdefinestyle{mathcode}{
  frame=single,
    language=Mathematica,
    backgroundcolor=\color{backcolour},
    commentstyle=\color{codegray},
    keywordstyle=\color{codeblue},
    numberstyle=\color{codegray}\ttfamily\tiny,
    stringstyle=\color{codegreen},
    basicstyle=\CODEstyle,
    breakatwhitespace=false,
    breaklines=true,
    postbreak=\mbox{\textcolor{codegray}{$\hookrightarrow$}\space},
    captionpos=b,
    keepspaces=true,
    numbers=left,
    numbersep=5pt,
    showspaces=false,
    showstringspaces=false,
    showtabs=false,
    tabsize=2,
    comment=[l]{\#},
}
\lstdefinestyle{shell}{
  frame=single,
	language=bash,
    commentstyle=\color{codegreen},
    keywordstyle=\color{codeblue},
    numberstyle=\color{codegray}\ttfamily\tiny,
    basicstyle=\CODEstyle,
    breakatwhitespace=false,
    breaklines=true,
    postbreak=\mbox{\textcolor{codegray}{$\hookrightarrow$}\space},
    captionpos=b,
    keepspaces=true,
    numbers=none,
    numbersep=5pt,
    showspaces=false,
    showstringspaces=false,
    showtabs=false,
    tabsize=2,
    comment=[l]{\#},
}

\makeatletter
\indexsetup{level=\relax,toclevel=section,noclearpage}%
\makeindex[name=cp,title=]%

{\catcode`\^^M=13 \gdef\GNUTobeylines{\catcode`\^^M=13 \def^^M{\null\par}}}

\setlist[description]{style=nextline, font=\normalfont}
\setlist[itemize]{label=\textbullet}
\makeatother

\newcounter{bla}

\journal{Computer Physics Communications}

\newpageafter{abstract}

\newcommand{\order}[1]{\ensuremath{{\cal O}(#1)}}
\newcommand{\deriv}[3]{\frac{\partial\ifthenelse{\equal{#1}{}}{}{^{#1}}%
    #2}{\partial #3\ifthenelse{\equal{#1}{}}{}{^{#1}}}}
\newcommand{\dderiv}[3]{\frac{\dd\ifthenelse{\equal{#1}{}}{}{^{#1}}%
    #2}{\dd #3\ifthenelse{\equal{#1}{}}{}{^{#1}}}}
\newcommand{\one}{one}
\newcommand{\two}{two}
\newcommand{\three}{three}
\newcommand{\ep}{\epsilon}
\newcommand{\dd}{\mathrm{d}}
\newcommand{\mathematica}{\code{Mathematica}}
\newcommand{\pysecdec}{\code{pySecDec}}
\newcommand{\pySecDec}{\pysecdec}
\newcommand{\ftint}{\code{ftint}}
\newcommand{\norm}[1]{(4\pi t)^{#1}}
\newcommand*{\abbrev}[1]{{\scalefont{.9}#1}}
\newcommand*{\citere}[1]{Ref.~\cite{#1}}
\newcommand*{\citeres}[1]{Refs.~\cite{#1}}
\newcommand*{\code}[1]{\texttt{#1}}

\newcommand*{\file}[1]{\texttt{#1}}

\newcommand*{\python}{\code{Python}}

\newcounter{notecount}

\newcommand{\myacrodef}[3]{\acrodef{#2}{#3}\newcommand{#1}{\ac{#2}}}
\myacrodef{\sm}{SM}{Standard Model}
\myacrodef{\qft}{QFT}{Quantum Field Theory}
\myacrodef{\gff}{GFF}{gradient-flow formalism}
\myacrodef{\sftx}{SFTX}{short-flow-time expansion}
\acrodefplural{QFT}{Quantum Field Theories}

\myacrodef{\QED}{QED}{Quantum Electrodynamics}
\myacrodef{\qcd}{QCD}{Quantum Chromodynamics}
\myacrodef{\eft}{EFT}{Effective Field Theory}
\acrodefplural{EFT}{Effective Field Theories}

\myacrodef{\smeft}{SMEFT}{Standard Model Effective Field Theory}
\myacrodef{\grsmeft}{GRSMEFT}{gravity-extension of \smeft}
\myacrodef{\LEFT}{LEFT}{Low-Energy Effective Field Theory}
\myacrodef{\wet}{WET}{Weak Effective Theory}
\myacrodef{\ibp}{IbP}{integration-by-parts}
\myacrodef{\eom}{EoM}{equation-of-motion}
\acrodefplural{EoM}{equations-of-motion}

\myacrodef{\mfv}{MFV}{Minimal Flavor Violation}
\myacrodef{\yd}{YD}{Young Diagram}
\acrodefplural{YD}[YDs]{Young Diagrams}

\myacrodef{\pypi}{PyPI}{Python Package Index}
\myacrodef{\uolea}{UOLEA}{Universal One-Loop Effective Action}

\usepackage{fancyhdr}
\newcommand{\RHheaderline}{\textsf{KA-TP-14-2024, TTK-24-28, P3H-24-050
--- July~2024}}  
\fancypagestyle{pprintTitle}
{
  
  \fancyhead[R]{\RHheaderline}
}
\VerbatimFootnotes

\begin{document}

\begin{frontmatter}

\title{\ftint: Calculating gradient-flow integrals with \pysecdec}

\author[1]{Robert V. Harlander}
\author[1]{Theodoros Nellopoulos}
\author[2]{Anton Olsson}
\author[3]{Marius Wesle}
\affiliation[1]{organization={TTK, RWTH Aachen University},
  city={52056 Aachen},
  country={Germany}}
\affiliation[2]{organization={Institute for Theoretical Physics, Karlsruhe
    Institute of Technology~(KIT)},
  city={76131~Karlsruhe},
  country={Germany}}
\affiliation[3]{organization={Department of Mathematics, University of
T\unexpanded{\"u}bingen},
city={72076~T{\"u}bingen},
country={Germany}}

\begin{abstract}
  The program \ftint\ is introduced which numerically evaluates dimensionally
  regularized integrals as they occur in the perturbative approach to the
  gradient-flow formalism in quantum field theory. It relies on sector
  decomposition in order to determine the coefficients of the individual
  orders in $\ep=(4-D)/2$, where $D$ is the space-time dimension. For that
  purpose, it implements an interface to the public library \pysecdec. The
  current version works for massive and massless integrals up to \three-loop
  level with vanishing external momenta, but the underlying method is
  extendable to more general cases.
\end{abstract}


\end{frontmatter}

\acresetall%

{\bf PROGRAM SUMMARY}

\begin{small}
\noindent
{\em Program title\/:}
\ftint{}

{\em Developer's repository link\/:}
\url{https://gitlab.com/ftint/ftint}

{\em Licensing provisions\/:}
MIT license (MIT)

{\em Programming language\/:}
\python{}

{\em Supplementary material\/:}
\file{README.md}

{\em Nature of problem\/:} The perturbative approach to the gradient-flow
formalism in quantum field theory leads to integrals which closely resemble
regular Feynman integrals. However, they involve exponential factors which
depend on the loop and external momenta, as well as on so-called flow-time
variables. In general, the latter are also integrated over a finite
interval. These integrals cannot be solved immediately with standard tools.

{\em Solution method\/:} The flow-time integrals are transformed to integrals
over a hypercube by introducing Schwinger parameters. The latter are
numerically evaluated using the public program \pysecdec, which performs a
sector decomposition and calculates the coefficients of the poles in the
parameter $\ep=(4-D)/2$ by numerical integration, where $D$ is the space-time
dimension which occurs in dimensional regularization.

{\em Additional comments including restrictions and unusual features\/:} In
its current form, \ftint\ is restricted to one-, two-, and three-loop
integrals with vanishing external momenta. It does allow for massive
propagators though, raised to virtually arbitrary integer powers.
\end{small}

\acresetall%

\clearpage
\tableofcontents
\clearpage

\section{Introduction}\label{sec:intro}

The \gff~\cite{Narayanan:2006rf,Luscher:2009eq,Luscher:2010iy,
  Luscher:2011bx,Luscher:2013cpa} is a useful tool for practical calculations
in \qft.  Its main feature is to suppress the high-momentum modes of quantum
fields. In lattice \qcd, this leads to a smoothing of the gauge field. Among
many other applications, this allows for efficient ways to determine the
lattice spacing, for example\,\cite{Luscher:2010iy,BMW:2012hcm}.

On the other hand, various applications of the \gff\ have been suggested that
involve also perturbative calculations, among them the so-called
\sftx~\cite{Luscher:2011bx}, where composite operators of flowed fields are
expressed in terms of regular operators via matching coefficients which can be
determined perturbatively. This approach has proven viable for evaluating
matrix elements of the energy-momentum tensor in \qcd, for example, or for the
calculation of observables in flavor
physics\,\cite{Suzuki:2013gza,Makino:2014taa,Iritani:2018idk,
  Harlander:2018zpi,Suzuki:2020zue,Suzuki:2021tlr,%
  Harlander:2022tgk,Black:2023vju}. For more applications, see
\citeres{Rizik:2020naq,Harlander:2020duo,Mereghetti:2021nkt,
  Harlander:2022vgf,Borgulat:2023xml,Shindler:2023xpd,Dragos:2019oxn}, for
example.

The form of the integrals that occur in the perturbative approach to the \gff\
is remarkably close to the Feynman integrals of regular \qft. The only
modifications are: (i)~an exponential factor in the integrand which depends on
the loop momenta, the external momenta and the masses, in general, as well as
on so-called flow-time variables, and (ii)~additional integrations over these
flow-time variables~\cite{Luscher:2010iy,Luscher:2011bx}. 

It has been shown that many of the tools that have been developed for
higher-order calculations in the perturbative approach to regular \qft\ can be
applied or extended to flowed \qft~\cite{Artz:2019bpr}. In particular, this
holds for the automatic generation of the associated Feynman diagrams, their
simplification to scalar integrals, as well as their reduction to master
integrals using \ibp\ relations. However, while there are powerful publicly
available software tools for the numerical evaluation of the master integrals
in regular \qft, this is not the case for flow-time integrals. In this paper,
we will close this gap by providing the program \ftint, which constitutes an
interface for such integrals
to \pySecDec~\cite{Borowka:2017idc,Borowka:2018goh,Heinrich:2023til},
allowing to evaluate them via the sector decomposition
algorithm~\cite{Binoth:2000ps,Binoth:2003ak,Heinrich:2008si}. In this first
version of the program we put the focus on integrals which depend only on a
single mass scale, the flow-time $t$. This is certainly one of the most
important cases, as the calculation of the matching coefficients in the \sftx\
leads to exactly this class of
integrals~\cite{Harlander:2018zpi}. However, \ftint\ also allows for
non-vanishing masses in the propagators, which may prove useful in many
applications such as the calculation of mass effects to the action density
$\langle G_{\mu\nu}G_{\mu\nu}\rangle$.

The structure of the remainder of this paper is as follows. After defining the
problem in \cref{sec:general}, we describe the method of our calculation
in \cref{sec:calculation}. The way to use the program \ftint\ is presented
in \cref{sec:ftint}, including some examples and
checks. \cref{sec:conclusions} contains conclusions and an outlook for
future work. In the appendix, we describe the possibility to adjust the input
and output format of \ftint.

\section{General outline}\label{sec:general}

We work in Euclidean space throughout this paper, unless stated otherwise.
The generic form of the integrals which we will consider
is~\cite{Harlander:2016vzb,Artz:2019bpr}
\begin{equation}\label{eq:cope}
  \begin{aligned}
    f(\mathbf{c},\mathbf{a},\mathbf{b}) &=
    \norm{lD/2}\,t^{-b}\int_{[0,1]^f} \dd
    \mathbf{u}\,\mathbf{u}^\mathbf{c} \int_{\mathbf{p}}
    \frac{\exp[-t\,\sum_{i=1}^k a_iP_i^2]}
         {(P_1^2+m_1^2)^{b_1}\cdots(P_k^2+m_k^2)^{b_k}}
  \end{aligned}
\end{equation}
where $\mathbf{p}=\{p_1,\ldots,p_l\}$ is the set of loop momenta,
\begin{equation}\label{eq:east}
  \begin{aligned}
    \mathbf{b} = \{\{b_1,m_1\},\ldots,\{b_k,m_k\}\}
  \end{aligned}
\end{equation}
collects the so-called indices $b_i\in\mathds{Z}$ and masses
$m_i\in\mathds{R}$. The prefactor $t^{-b}$, with $b=\sum_{i=1}^k b_i$, in
\cref{eq:cope} thus compensates the mass dimension of the integral, such that
$f(\mathbf{c},\mathbf{a},\mathbf{b})$ is dimensionless. We furthermore define
the simplified notation $\{b_i,0\}\to b_i$ for massless propagators.

The $k$ functions $a_i\in\mathbf{a}$ are real-valued polynomials of the
(dimensionless) flow-time variables $\mathbf{u}=\{u_1,\ldots,u_f\}$ that are
non-negative on the hypercube $[0,1]^f$.  Furthermore, we define
\begin{equation}\label{eq:cote}
  \begin{aligned}
    \mathbf{u}^\mathbf{c} \coloneq u_1^{c_1}\cdots u_f^{c_f}\,,
  \end{aligned}
\end{equation}
where $\mathbf{c}=\{c_1,\ldots,c_f\}$ is a set of non-negative integers.

The integration measure over the loop momenta is
\begin{equation}\label{eq:dort}
  \begin{aligned}
    \int_\mathbf{p} \coloneq \int_{p_1}\cdots \int_{p_l}\,,\quad\mbox{with}\quad
    \int_p=\int\frac{\dd^D p}{(2\pi)^D}\,,\qquad D= 4-2\ep\,.
  \end{aligned}
\end{equation}
The $P_i$ in \cref{eq:cope} are linear combinations of the loop
momenta. External momenta will be neglected in this paper. This is justified
for the calculation of the perturbative matching coefficients of the \sftx\ to
which \ftint\ is tailored.\footnote{The matching coefficients are most
conveniently determined by using the method of
projectors~\cite{Gorishnii:1983su,Gorishnii:1986gn}, which results in
integrals whose only dimensional scale is the external flow-time $t$,
see \citere{Harlander:2018zpi}, for example. } It implies that the number of
propagators $k$ in \cref{eq:cope} is related to the number of loops $l$ as
\begin{equation}\label{eq:hims}
  \begin{aligned}
    k=\left\{\begin{array}{lll}
    1 & \mbox{for} & l=1\,,\\
    3\,(l-1) & \mbox{for} & l\geq 2\,.
    \end{array}
    \right.
  \end{aligned}
\end{equation}
In principle, there could be propagator factors in \cref{eq:cope} which
differ only in their mass term, i.e., $P_i=P_j$ but $m_i\neq m_j$. However,
using partial fractioning, these can always be re-written to integrals where
$m_i=m_j$ if $P_i=P_j$.

\begin{figure}
  \begin{center}
    \begin{tabular}{ccc}
      \raisebox{-.2em}{%
        \mbox{%
          \includegraphics[%
            width=.3\textwidth]%
                          {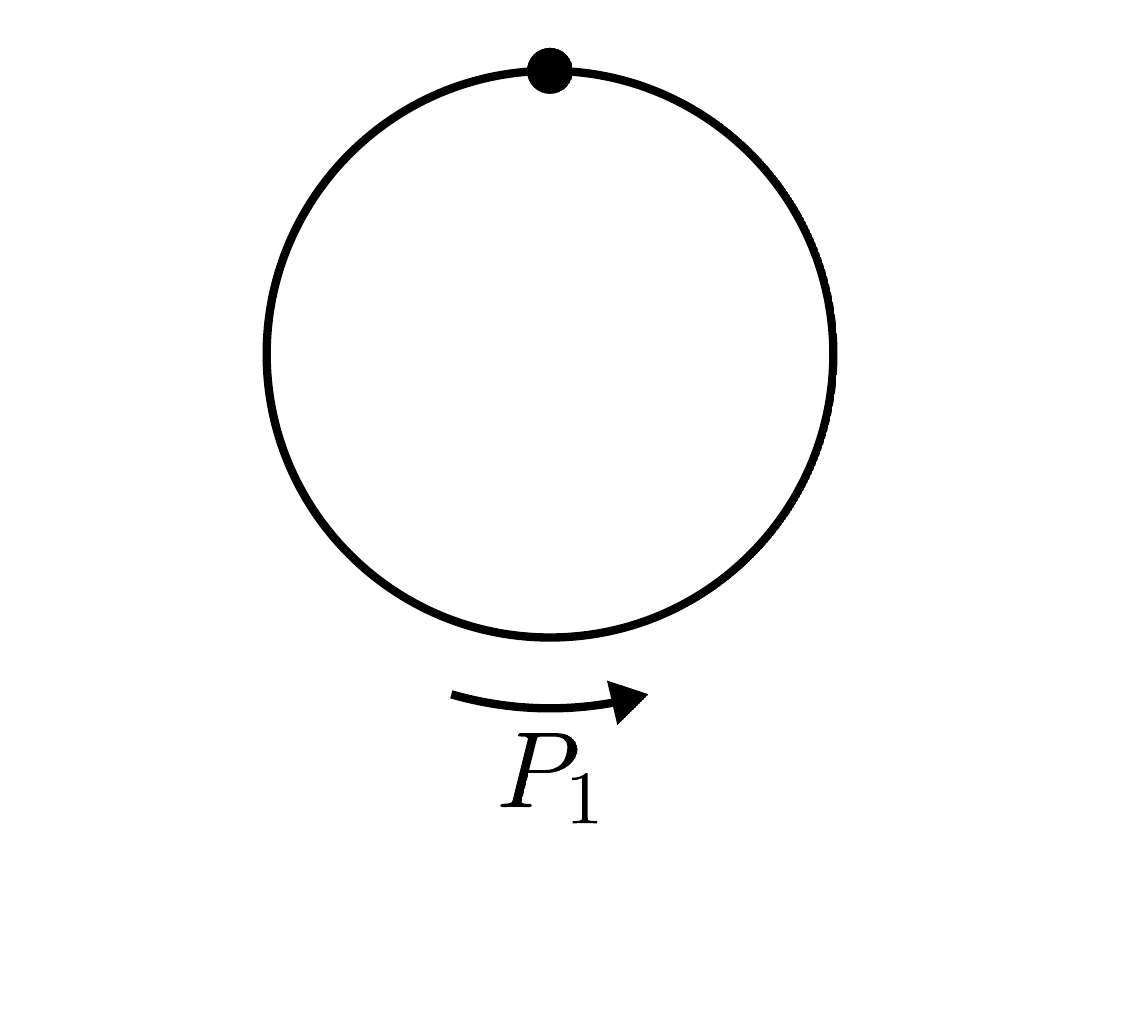}}} &
      \raisebox{.5em}{%
        \mbox{%
          \includegraphics[%
            width=.3\textwidth]%
                          {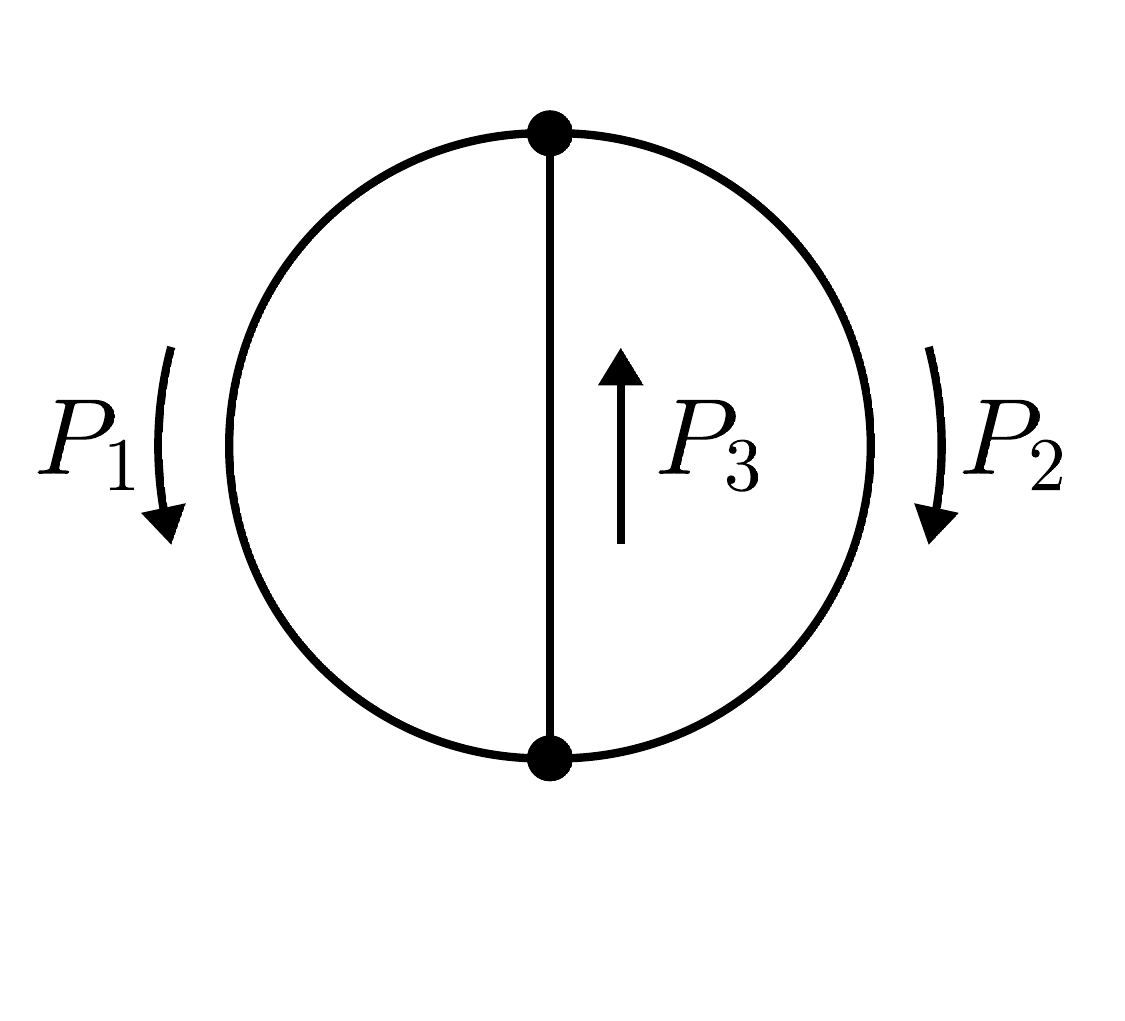}}} &
      \raisebox{0em}{%
        \mbox{%
          \includegraphics[%
            width=.3\textwidth]%
                          {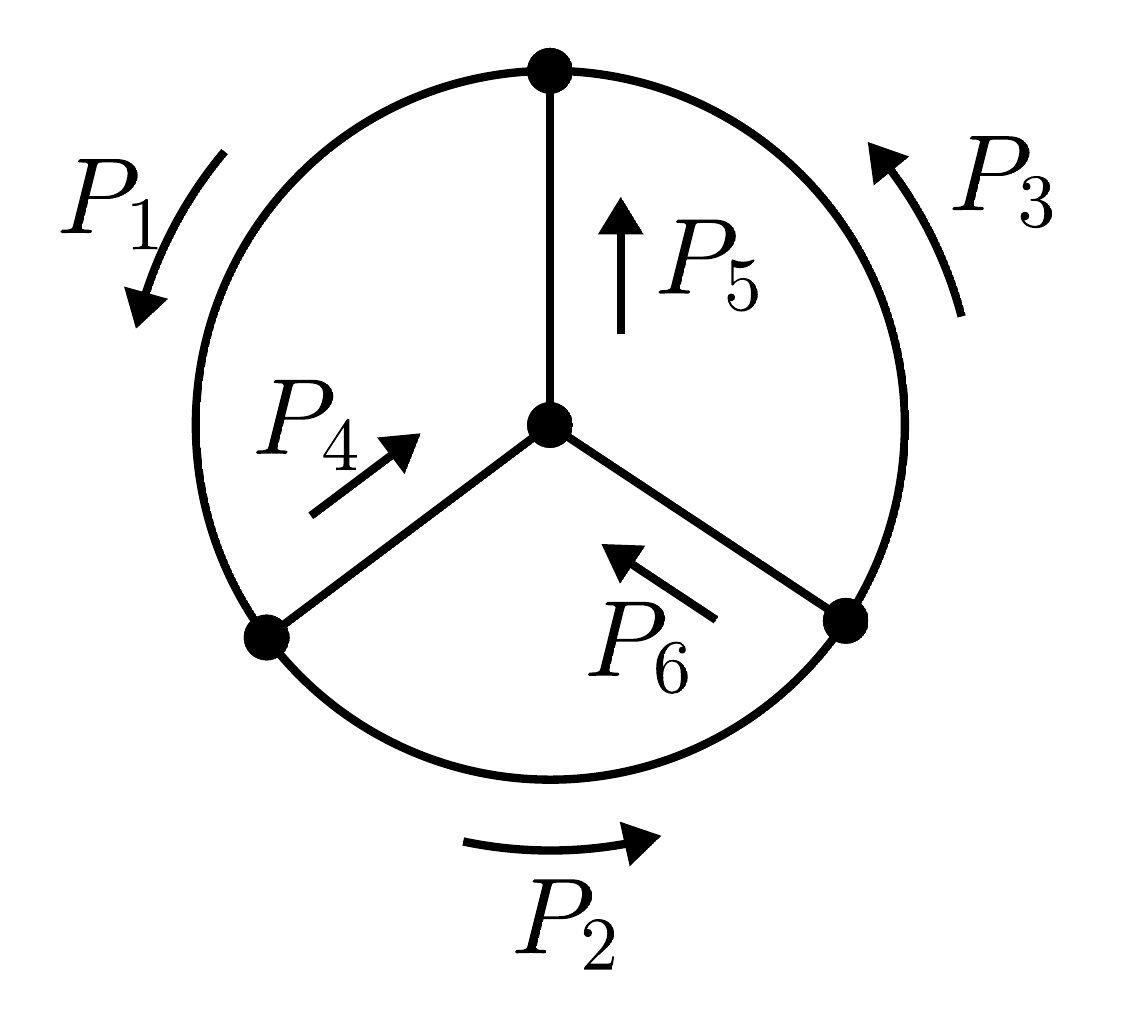}}}
    \end{tabular}
    \parbox{.9\textwidth}{
      \caption[]{\label{fig:topologies}\sloppy One-, two-, and three-loop
        topologies of the integrals considered in this paper. All Feynman
    diagrams are produced with the help
    of \code{FeynGame}\,\cite{Harlander:2020cyh,Harlander:2024qbn}.}
    }
  \end{center}
\end{figure}

For later purposes, it is convenient to define a symmetric square matrix
$A(\mathbf{a})$ through the condition
\begin{equation}\label{eq:bass}
  \begin{aligned}
    \mathbf{p}^T\,A(\mathbf{a})\mathbf{p} = \sum_{i=1}^k a_i P^2_i\,,
  \end{aligned}
\end{equation}
where $\mathbf{p}$ are the loop momenta.  Its explicit form depends on the
choice of linear combinations $P_i$. To be specific, we choose
\begin{description}
\item[\one-loop:]
  \begin{subequations}
  \begin{equation}\label{eq:deli1}
    \begin{aligned}
      P_1 &= p_1\,,\\
      \Rightarrow\      A(\mathbf{a}) &= a_1\,.
    \end{aligned}
  \end{equation}
  \end{subequations}
\item[\two-loop:]
    \begin{subequations}\label{eq:deli2}
      \begin{equation}
    \begin{aligned}
      P_1 &= p_1\,,\qquad
      P_2 = p_2\,,\qquad
      P_3 = p_1+p_2\,,
    \end{aligned}
  \end{equation}
  \begin{equation}
    \begin{aligned}
    \Rightarrow\     A(\mathbf{a})&=\left(
    \begin{matrix}
      a_1+a_3 & a_3\\
      a_3 & a_2+a_3
    \end{matrix}
    \right)\,.\\
    \end{aligned}
  \end{equation}
    \end{subequations}
\item[\three-loop:]
    \begin{subequations}
  \begin{equation}\label{eq:deli3a}
    \begin{aligned}
      P_1 &= p_1\,,&\qquad
      P_2 &= p_2\,,&\qquad
      P_3 &= p_3\,,\\
      P_4 &= p_1-p_2\,,&\qquad
      P_5 &= p_1-p_3\,,&\qquad
      P_6 &= p_2-p_3\,,\\
    \end{aligned}
  \end{equation}
  \begin{equation}\label{eq:deli3b}
    \begin{aligned}
    \Rightarrow\    A(\mathbf{a}) &=
    \left(
    \begin{matrix}
      a_1 + a_4 + a_5 & - a_4 & - a_5 \\
      -a_4 & a_2 + a_4 + a_6 & -a_6 \\
      -a_5 & -a_6 & a_3 + a_5 + a_6
    \end{matrix}
    \right)\,.
    \end{aligned}
  \end{equation}
    \end{subequations}
\end{description}
This choice defines the so-called integral topologies shown in
\cref{fig:topologies}. Note that there is only a single topology at each loop
order up to the three-loop level. This pattern does not extend to higher
orders though.

\section{Calculation of the integrals}\label{sec:calculation}

The method we pursue for the evaluation of the integrals closely
follows \citere{Harlander:2016vzb}. However, while this paper employed the
public software tools \code{FIESTA}~\cite{Smirnov:2008py,Smirnov:2009pb,
Smirnov:2013eza,Smirnov:2015mct} and the \code{MPFR} integration
library,\footnote{See \url{http://www.holoborodko.com/pavel/mpfr/} and
\url{https://www.mpfr.org/}} we use \pySecDec\ for the main calculational steps.

\subsection{One loop}

In order to describe the method that we apply for the calculation of the
integrals, it is instructive to consider the one-loop level first. This is a
particularly simple case which, however, already introduces the main
concepts. The formal step to a general number of loops can then be achieved
rather easily.  Furthermore, it is useful to neglect all masses in a first
outline. It turns out that their inclusion is almost trivial in the approach
described below.

\subsubsection{Massless case}

At the one-loop level and neglecting the masses, \cref{eq:cope} takes the
form
\begin{equation}\label{eq:arri}
  \begin{aligned}
    f(\mathbf{c},\{a\},\{b\}) &= \norm{D/2}\,t^{-b}\int_{[0,1]^f} \dd
    \mathbf{u}\,\mathbf{u}^\mathbf{c} \int_p
    \frac{\exp[-tap^2]}{(p^2)^b}\,.
  \end{aligned}
\end{equation}
It is useful to distinguish three cases.
\begin{description}
\item{\textbf{Vanishing index.}}
For $b=0$, evaluating the integral over $p$ gives
\begin{equation}\label{eq:adie}
  \begin{aligned}
    f(\mathbf{c},\{a\},\{0\}) &= 
    \int_{[0,1]^f} \dd \mathbf{u}\,\mathbf{u}^\mathbf{c}\,
        [a]^{-D/2}\,,
  \end{aligned}
\end{equation}
Recall that $a$ is a polynomial in the flow-time variables $u_i$.  This can be
passed directly to \pysecdec\ for numerical integration, see
\cref{sec:pysecdec}.
\item{\textbf{Positive index.}} For $b>0$, we apply Schwinger parameters to
make the $p$ integral Gau\ss{}ian:
\begin{equation}\label{eq:jael}
  \begin{aligned}
    f(\mathbf{c}&,\{a\},\{b\}) = \frac{\norm{D/2}}{(b-1)!}
    \int_0^\infty\dd
    x\,x^{b-1} \int_{[0,1]^f} \dd \mathbf{u}\,\mathbf{u}^\mathbf{c} \int_p
    \exp\{-t[a+x]p^2\} =\\ &= 
    \frac{1}{(b-1)!}\int_0^\infty\dd x\,x^{b-1} \int_{[0,1]^f} \dd
    \mathbf{u}\,\mathbf{u}^\mathbf{c}\, [a+x]^{-D/2}\,.
  \end{aligned}
\end{equation}
In order to evaluate the integral over Schwinger parameters with \pysecdec, we
map them to the interval $[0,1]$. Unfortunately, the mapping
\begin{equation}\label{eq:hose}
  \begin{aligned}
    x\to\frac{1-x}{x}
  \end{aligned}
\end{equation}
introduces singularities at $x=1$, while \pysecdec{} expects them to
occur only at $x=0$. Therefore, we split the integration intervals of the
Schwinger parameters as $[0,\infty) = [0,1]\cup (1,\infty)$, and map
$x\to 1/x$ in the second interval, leading to
\begin{equation}\label{eq:bara}
  \begin{aligned}
    f(\mathbf{c}&,\{a\},\{b\}) = 
    \frac{1}{(b-1)!}
    \int_{[0,1]^f} \dd \mathbf{u}\,\mathbf{u}^\mathbf{c}\\&
    \times\int_0^1\dd x
    \left(
        x^{b-1}\,[a+x]^{-D/2}
        + x^{D/2-1-b}\,[x\,a+1]^{-D/2}\right)\,.
  \end{aligned}
\end{equation}
\pysecdec{} can now treat the $x$ and $\mathbf{u}$ integration on the
same footing.
\item{\textbf{Negative index.}}
In the case $b<0$, we can use
\begin{equation}\label{eq:huge}
  \begin{aligned}
    (p^2)^{-b} = (-t)^b\deriv{-b}{}{x}e^{-xtp^2}\bigg|_{x=0}\,
  \end{aligned}
\end{equation}
and obtain
\begin{equation}\label{eq:deli}
  \begin{aligned}
  f(\mathbf{c},\{a\},\{b\})
  &= (-1)^b\deriv{-b}{}{x}\int_{[0,1]^f} \dd \mathbf{u}\,\mathbf{u}^\mathbf{c}
                                     [a + x]^{-D/2}\bigg|_{x=0}\\
  &= P(-b)\,(-1)^b\int_{[0,1]^f} \dd \mathbf{u}\,\mathbf{u}^\mathbf{c}
                                     [a]^{-D/2+b}\,,
    \end{aligned}
\end{equation}
where
\begin{equation}\label{eq:gate}
  \begin{aligned}
    P(m)= \prod_{k=1}^{m}(1-k-D/2) = \frac{\Gamma(1-D/2)}{\Gamma(1-D/2-m)}\,.
  \end{aligned}
\end{equation}
This expression can again be directly passed to \pysecdec.
\end{description}

Let us look at a few simple examples which can be calculated analytically:
\begin{equation}\label{eq:farl}
  \begin{aligned}
    f(\{\},\{1\},&\{0\}) = 1\,,\\
    f(\{\},\{a\},&\{n\}) = \norm{D/2}\,t^{-n}\int_p
    \frac{\exp[-tap^2]}{(p^2)^n}
    =\\&=\frac{a^{n-D/2}}{(n-1)!}\int_0^\infty\dd x\,x^{n-1}
    (1+x)^{-D/2}=\\&
    =\frac{a^{n-D/2}}{(n-1)!}\frac{\Gamma(n)\Gamma(D/2-n)}{\Gamma(D/2)}\,,
    \qquad\mbox{for}\quad n\geq1\,,a\in\mathds{R}\,,
    \\
    f(\{\},\{a\},&\{-n\})
    = \norm{D/2}\,t^n\int_p\,p^{2n}\exp[-tap^2] =\\&=
    \norm{D/2}(-a)^{-n}\deriv{n}{}{x}\int_p\exp[-ta(1+x)p^2]\bigg|_{x=0} =
    \\ &=(-1)^{n}a^{-n-D/2}\deriv{n}{}{x}(1+x)^{-D/2}\bigg|_{x=0}\,,
  \end{aligned}
\end{equation}
from which it follows that
\begin{equation}\label{eq:hire}
  \begin{aligned}
    f(\{\},\{1\},&\{1\})
    =\frac{1}{D/2-1} =
    1+\ep+\ep^2+\cdots\,,\\
    f(\{\},\{1\},&\{2\}) =
    -\frac{1}{\ep}\left(1+\ep+\ep^2+\ldots\right)\,,\\
    f(\{0\},\{u_1\},&\{1\}) =
    \norm{D/2}\,t^{-1}\int_0^1\dd u_1\int_p
    \frac{\exp[-tu_1p^2]}{p^2} =\\&
    =-\norm{D/2}\,t^{-2}\int_p
    \frac{\exp[-tp^2]}{(p^2)^2} 
    =-f(\{0\},\{1\},\{2\})\,,\\
    f(\{0\},\{u_1\},&\{-1\}) = -1\,,
  \end{aligned}
\end{equation}
for example, where we have used the fact that scaleless integrals are zero in
dimensional regularization.

\subsubsection{Massive propagators}\label{sec:massive_propagators}

We may generalize the integral in \cref{eq:arri} by allowing for massive
propagators. Specifically, we consider the integral
\begin{equation}\label{eq:arrim}
  \begin{aligned}
    f(\mathbf{c},\{a\},\{\{b,m\}\}) &\equiv
    \norm{D/2}\,t^{-b}\int_{[0,1]^f} \dd \mathbf{u}\,\mathbf{u}^\mathbf{c}
    \int_p
    \frac{\exp[-tap^2]}{(p^2+m^2)^{b}}\,.
  \end{aligned}
\end{equation}
Note that we only need to consider positive $b$, because the other cases can
be algebraically reduced to already known integrals via the binomial formula:
\begin{equation}\label{eq:baku}
  \begin{aligned}
    (p^2+m^2)^{-b} = \sum_{n=0}^{-b}
    \left(
    \begin{matrix}
      -b\\n
    \end{matrix}
    \right)(p^2)^n(m^2)^{-b-n}\qquad\text{for}\ b\leq 0\,.
  \end{aligned}
\end{equation}

Following the same steps as above for $b>0$, on the other hand, we arrive at
\begin{equation}\label{eq:jaelm}
  \begin{aligned}
    f&(\mathbf{c},\{a\},\{\{b,m\}\}) =\\&= 
    \frac{1}{(b-1)!}\int_0^\infty\dd x\,x^{b-1} e^{-xtm^2}\int_{[0,1]^f} \dd
    \mathbf{u}\,\mathbf{u}^\mathbf{c}\, [a+x]^{-D/2}\,.
  \end{aligned}
\end{equation}
We again split the integration interval into $x\in [0,1] \cup (1,\infty)$ and
perform the substitution $x\to 1/x$ in the second interval. This leads to a
singularity at $x=0$ in the argument of the exponential, which is spurious,
however, as the exponential vanishes at this point. Therefore, the only effect
on the integrals due to propagators being massive is the multiplication of a
positive function. Since it is factorized and does not contribute to the true
singularity structure, it has no impact on pole extraction during sector
decomposition. \cref{sec:pysecdec} describes how such factors are treated in
\pysecdec.

As mentioned above, the more general case of several masses,
\begin{equation}\label{eq:jaelm1}
  \begin{aligned}
    \int_{[0,1]^f} \dd \mathbf{u}\,\mathbf{u}^\mathbf{c}
    \int_p
    \frac{\exp[-tap^2]}{
      (p^2+m_1^2)^{b_1}
      \cdots
      (p^2+m_n^2)^{b_n}
    }
  \end{aligned}
\end{equation}
can be reduced to integrals of the form \cref{eq:jaelm} by partial
fractioning.

Let us consider a particularly simple example for a massive one-loop integral
which can be solved analytically:
\begin{equation}\label{eq:massive:gens}
  \begin{aligned}
    f(\{\},\{1\},\{\{1,1\}\}) = 1-z e^{z}\Gamma(0,z)\,,\qquad z=m^2t\,,
  \end{aligned}
\end{equation}
where $\Gamma(n,z)=\int_{z}^\infty\dd x\,x^{n-1}e^{-x}$ is the incomplete
$\Gamma$ function.

\subsection{Higher orders}\label{sec:2loop}

Let us now move on to the multi-loop level, first focusing on the massless
case. It is helpful to define the auxiliary function
\begin{equation}\label{eq:hsia}
  \begin{aligned}
    F(\mathbf{a},\mathbf{x}) =
    \norm{lD/2}        \int_\mathbf{p}
        \exp\left\{-t\sum_{i=1}^k [a_i+x_i] P_i^2\right\}\,.
  \end{aligned}
\end{equation}
Using \cref{eq:bass}, one can perform the Gau\ss{}ian integral over the loop
momenta to obtain
\begin{equation}\label{eq:amis}
  \begin{aligned}
    F(\mathbf{a},\mathbf{x}) =[\det
      A(\mathbf{a}+\mathbf{x})]^{-D/2}\,.
  \end{aligned}
\end{equation}
We refer to the $\mathbf{x}=\{x_1,\ldots,x_k\}$ as Schwinger parameters in the
following, even if they are not integrated over. It is helpful to note that
each $x_i\in \mathbf{x}$ occurs only linearly in $\det
A(\mathbf{a}+\mathbf{x})$.

Consider now a flow-time integral with indices $b_1,\ldots,b_k$, which we
divide up as follows:
\begin{equation}\label{eq:cara}
  \begin{aligned}
    b_i &> 0 \quad\mbox{for}\quad i\in I_\text{int}\,,\\
    b_i &< 0 \quad\mbox{for}\quad i\in I_\text{diff}\,,\\
    b_i &= 0 \quad\mbox{for}\quad i\in I_0\,.
  \end{aligned}
\end{equation}
For the vanishing indices $b_i=0$, we can simply set the corresponding
Schwinger parameters to zero in \cref{eq:amis}, $x_i=0$. For the negative
indices $b_i<0$, on the other hand, we use \cref{eq:huge}, meaning that we
need to take the derivative w.r.t.\ $-x_i$ at $x_i=0$. Note that $n$
derivatives acting on \cref{eq:amis} produce $n$ terms of the form
\begin{equation}\label{eq:isbn}
  \begin{aligned}
    [\det A(\mathbf{a}+\mathbf{x})]^{-D/2-k}\,g_k(\mathbf{a},\mathbf{x},D)\,,
    \qquad k\in\{1,\ldots,n\}\,,
  \end{aligned}
\end{equation}
where $g_k$ is polynomial in its arguments at most of order $x_i^k$ for each
$x_i$.  Finally, for the positive indices $b_i>0$, we integrate over $x_i$ and
multiply by $1/(b_i-1)!$.  In summary,
\begin{equation}\label{eq:cham}
  \begin{aligned}
    f(\mathbf{c},\mathbf{a},\mathbf{b})
       &=
    \left[\prod_{j\in I_\text{int}}\frac{1}{(b_j-1)!}
          \int_0^\infty\dd x_j\, x_j^{b_j-1}\right]\times\\&
    \hspace*{-3em}\times
    \int_{[0,1]^f} \dd
    \mathbf{u}\,\mathbf{u}^\mathbf{c}\,
            \left[\left(\prod_{i\in I_\text{diff}}
    \deriv{-b_i}{}{(-x_i)}\right) [\det
      A(\mathbf{a}+\mathbf{x})]^{-D/2}\right]_{x_k=0\ \mbox{for}\ 
              k\in (I_\text{diff}\cup I_0)}\,.
  \end{aligned}
\end{equation}
Again, we split the integration region for the Schwinger parameters into
$x_j\in [0,1]\cup (1,\infty]$.  The integrand in \cref{eq:cham} thus consists
of polynomials of the $x_i$ with $i\in I_\text{int}$ and the $\mathbf{u}$,
raised to non-integer powers which can be passed to \pysecdec\ for
integration.

Non-vanishing masses $\mathbf{m}$ can be taken into account in a
straight-forward way. As pointed out above, only massive propagators with
positive indices need to be considered, because non-positive indices can be
reduced algebraically to known integrals. Also, we can assume that each
independent momentum is associated with only a single mass, which can always
be achieved by partial fractioning as pointed out in the one-loop case.  In
this case, the only modification is to include a factor
\begin{equation}\label{eq:iowa}
  \begin{aligned}
    \exp\left(-t\sum_{j\in I_\mathrm{int}} x_jm_j^2\right)
  \end{aligned}
\end{equation}
in the integrand of \cref{eq:cham}.

\subsection{Symmetries of the flow-time integrals}\label{sec:normalform}

The representation of a flow-time integral in the form of \cref{eq:cope} is
not unique. The integrals remain invariant under certain combined permutations
of the parameters $\mathbf{c}$, $\mathbf{a}$, $\mathbf{u}$, and $\mathbf{b}$.
Employing such symmetries may significantly reduce the number of integrals
that need to be evaluated. \ftint\ provides an option to map any flow-time
integral to a standard form which we refer to as \textit{normal form}. This
section briefly describes our basic strategy to determine the normal form.

The first symmetry we employ corresponds to re-naming the flow-time
integration variables. In general, the integral is preserved when permuting
the variables $\mathbf{u}$ in the polynomials $\mathbf{a}(\mathbf{u})$ and
applying the inverse permutation to $\mathbf{c}$, i.e.
\begin{equation}\label{eq:normalform:bloc}
  \begin{aligned}
    f(\mathbf{c},\mathbf{a}(\mathbf{u}),\mathbf{b})
    = f(p\mathbf{c},\mathbf{a}(p^{-1}\mathbf{u}),\mathbf{b})\,,
  \end{aligned}
\end{equation}
where $p$ denotes a permutation and $p^{-1}$ its inverse.  For example, one
arrives at the identity
\begin{equation}\label{eq:rename_flow-time}
  \begin{aligned}
    f(\{0,1\},\{u_1u_2,u_2,u_1\},\{3,1,2\}) 
    = f(\{1,0\},\{u_1u_2,u_1,u_2\},\{3,1,2\})
  \end{aligned}
\end{equation} 
by just interchanging the names of $u_1$ and $u_2$.

The second symmetry is related to permutations of the momenta $P_i$ (modulo
signs) which leave the topologies in \cref{fig:topologies} (i.e., the
momentum conservation relations) invariant. Since each line of these
topologies corresponds to a momentum $P_i$, an index $b_i$ and a polynomial
$a_i$, permutations of the $P_i$ correspond to simultaneous permutations of
the $b_i$ and the $a_i$.  In the one-loop case, there is only a single line
and thus no additional symmetry results from these considerations. The
two-loop topology, on the other hand, is symmetric under any permutation of
lines, and thus any simultaneous permutations of $\mathbf{a}$ and
$\mathbf{b}$. For example,
\begin{equation}\label{eq:rename_momentum}
  \begin{aligned}
    f(\{1,0\},\{u_1u_2,u_1,u_2\},\{3,1,2\})
    =f(\{1,0\},\{u_1u_2,u_2,u_1\},\{3,2,1\})\,. 
  \end{aligned}
\end{equation}
At three-loop level, there are $4!=24$ permutations of the $P_i$ that preserve
the topology shown in \cref{fig:topologies} (corresponding to the permutations
of the four vertices). For example,
\begin{equation}\label{eq:normalform:irja}
  \begin{aligned}
    (P_1,P_2,P_3,P_4,P_5,P_6)\to
    (-P_3,-P_2,-P_1,P_6,P_5,P_4)
  \end{aligned}
\end{equation}
corresponds to the mirror symmetry along the vertical axis, while
\begin{equation}\label{eq:normalform:irja1}
  \begin{aligned}
    (P_1,P_2,P_3,P_4,P_5,P_6)\to
    (-P_3,P_6,P_5,-P_2,-P_1,-P_4)
  \end{aligned}
\end{equation}
implies a non-trivial continuous deformation of the diagram. Each such
transformation results in a permutation $p$, for which it holds
that\footnote{Since only the squares of the $P_i$ enter the integral, sign
changes of the $P_i$ like in \cref{eq:normalform:irja,eq:normalform:irja1} do
not matter here.}
\begin{equation}\label{eq:normalform:bloc1}
  \begin{aligned}
    f(\mathbf{c},\mathbf{a},\mathbf{b})
    = f(\mathbf{c},p\mathbf{a}, p\mathbf{b})\,.
  \end{aligned}
\end{equation}

Using these symmetries, we can map any flow-time integral $I$ onto an
equivalent standard form. To achieve this, we generate a list of equivalent
integrals by applying all combinations of permutations from each of the two
symmetries to $I$. This list then gets sorted according to a lexicographical
criterion, and the first element is defined to be the normal form of $I$. Our
procedure guarantees that two flow-time integrals can be transformed into one
another by the symmetry operations described above if and only if they have
the same normal form; they will then obviously integrate to the same result.
Note that the converse is not true: Integrals with different normal forms
cannot be transformed into one another by the discussed symmetry operations;
however, they may still integrate to the same result.

A specific example and how to use \ftint\ in order to map an integral to its
normal form will be discussed in \cref{sec:normalform_map}.

\subsection{Implementation with \pysecdec}\label{sec:pysecdec}
\pysecdec\ \cite{Borowka:2017idc,Borowka:2018goh,
Heinrich:2023til} is a toolbox for the evaluation of dimensionally regularized
parameter integrals. It utilizes the sector decomposition algorithm to isolate
and subtract overlapping endpoint singularities, and produces an integration
library to evaluate the coefficients of an expansion in the dimensional
regulator. The parameter integrals \pysecdec\ targets are of the form
\begin{equation}
  \label{eq:parameter_integral}
  I = \int_{[0,1]^d} \text{d}\mathbf{x} \; f_1^{\alpha_1}(\mathbf{x})\cdots
  f_k^{\alpha_k}(\mathbf{x}),
\end{equation}
where the $f_l$ are functions of the parameters and the $\alpha_l$ are 
linear in the space-time dimension $D$. These integrals are divergent
in general, but can be evaluated in dimensional regularization by taking 
$D=4-2\ep$ and extracting the poles in a Laurent series in $\ep$. Performing
such an expansion requires defining adequate subtraction terms, which can be 
highly non-trivial for integrands with nested singularity structures.
The sector decomposition approach offers an algorithmic procedure of decomposing
the integral into sectors with factorized singularity structures, where it is
straightforward to define subtractions 
\cite{Binoth:2000ps,Binoth:2003ak,Heinrich:2008si}.
\pysecdec\ provides implementations of several decomposition algorithms,
based on either iterative or geometric strategies. The interface in \ftint\
uses geometric sector decomposition as it usually leads to fewer sectors than
the iterative approaches \cite{Kaneko:2009qx,Heinrich:2021dbf}. After sector
decomposition, the integral is represented as a sum of Laurent series in
$\alpha$ sectors
\begin{equation}
  \label{eq:decomposed_integral}
  I = \sum_{l=1}^{\alpha} \sum_{n=-r}^{p} I_{l,n} 
        \frac{1}{\ep^n} +\mathcal{O}(\ep^{r+1}),
\end{equation}
where $p$ is the order of the highest pole and the expansion coefficients 
$I_{l,n}$ are sector integrals that are finite at the integration boundaries.
A simple example of a sector integral $I_{j,0}$ making up the
finite part of the expansion in a sector $j$ where a logarithmic divergence 
in $x_1$ has been extracted is
\begin{equation}
  \label{eq:sector_integral}
  I_{j,0} = \int_{[0,1]^d} \text{d}\mathbf{x} \; x_1^{-1-\ep} 
  \left[\mathcal{I}(x_1,\dots,x_d)-\mathcal{I}(x_1=0,\dots,x_d)\right].
\end{equation}
The subtraction term $\mathcal{I}(x_1=0,\dots,x_d)$ ensures that the sector
integral is finite as $x_1 \rightarrow 0$. For more severe divergences, the
power of the extracted pole is raised through a number of integration-by-parts
iterations until there are only logarithmic divergences remaining.  In this
form, the sector integrals are well suited for numerical integration. The
latest release of \pysecdec\ \cite{Heinrich:2023til}
introduced \code{Disteval}, a new integration library. It implements a
quasi-Monte Carlo integrator which has yielded significant performance
increases compared to previous versions. The integration interface in \ftint\
exclusively uses the \code{Disteval} integrator as it supersedes all older
integrators.

The massive flow-time integrals described in \cref{sec:massive_propagators}
include factors $e^{-tm_j^2x_j}$. After the substitution $x\to 1/x$ in the
second interval (to map $(1,\infty) \to (0,1)$) they transform into
$e^{-tm_j^2/x_j}$. The $1/x$ pole in the argument is spurious and
does not need to be extracted through sector decomposition. To
minimize the work of the decomposition algorithm, \pysecdec\ allows for the 
definition of finite functions that only enter at subtraction level. 
Factors like these effectively scale the magnitude of the integrand, 
which means they still need to be included in subtraction terms such as 
$\mathcal{I}(x_1=0,\dots,x_d)$ in \cref{eq:sector_integral}. In order to
avoid practical issues of having a spurious $1/x$ pole in the exponent, a
regulator $\delta$ is added to the transformed exponential functions such that
\begin{equation}\label{eq:regulated_exponentials}
  e^{-tm_j^2/x_j} \to e^{-tm_j^2/(x_j+\delta)}.
\end{equation}
By default, $\delta$ is set to $10^{-10}$ at integration.  In order to ensure
that this parameter does not affect the accuracy of the integration result,
the user may change it using the option \code{---delta},
see \cref{sec:numint}.  Since the exponential function does not affect sector
decomposition, this additional parameter has virtually no impact on
performance.

\section{Using \ftint}\label{sec:ftint}

Following the spirit of \pySecDec, the evaluation of an integral with \ftint\
is divided into two steps.  In the first step, the integral is decomposed into
sectors where the boundary singularities have been isolated and subtracted
with the help of \pysecdec. It creates and compiles a
\code{C++} integration library which is used in the second step to
numerically evaluate the integrals. The motivation for splitting the program
into two parts is that the user may want to perform the second step several
times, for example with different target errors or for several mass
parameters. Since this only affects the numerical integration, one can save on
computing time for the decomposition and compilation step which needs to be
done only once in this case.

It is important to note that the current version of \ftint\ is limited to
evaluating integrals of the form specified in \cref{eq:cope}. Specifically,
this implies that all external momenta must be zero, all massive propagators
must have a positive index $b$, and no two propagators may carry the same
momentum while having different masses. As previously mentioned, the latter
two constraints do not really represent limitations, as any integral that does
not meet these criteria can be algebraically transformed into a linear
combination of integrals that do. While future versions of \ftint\ may include
built-in support for these transformations, the current version requires the
user to apply these modifications before passing them to \ftint. This can be
readily accomplished using built-in functions
in \mathematica~\cite{Mathematica}
or \code{FORM}~\cite{Vermaseren:2000nd,Kuipers:2012rf}, for example, or via
the \python\ library \code{SymPy}.

\subsection{Sector decomposition}

The decomposition part of the program is implemented in
\code{ftint\_pySecDec.py}.  It is called from the command line as
\begin{lstlisting}[style=shell,
        caption={Command for sector decomposition.},
        label={lst:deccall}]
  $ python3 <ftint_path>/ftint_pySecDec.py <ft_integrals> [<options>]
\end{lstlisting}
where, here and in the following, \code{<ftint\_path>} is to be replaced by the
actual path to the source code of \ftint.  In the simplest case,
\code{<ft\_integrals>} is a string encoding a single flow-time integral. Its
format follows closely the definition in \cref{eq:cope}, with some
adjustments. The most significant one is the representation of the masses. For
the sector decomposition, the actual value of a mass is irrelevant, as long as
it is non-zero. In the input for \code{ftint\_pySecDec.py}, non-zero masses are
indicated only by their index, i.e.\ $m_1\to \verb/1/$, $m_2\to \verb/2/$,
etc.\footnote{The reason why we do not adopt the more intuitive
notation \verb$f[{0,1},{u2-u1*u2,2,u2},{-1,{2,m2},{1,m3}}]$ is not to
interfere with any locally defined symbols \code{m2}, \code{m3} in the
user's code.} Let us consider a specific \two-loop
example:\footnote{See \code{examples/2L\_massive.in} in the code repository.}
\begin{equation}\label{eq:baor}
  \begin{aligned}
& \verb$f[{0,1},{u2-u1*u2,2,u2},{-1,{2,2},{1,3}}]$ =\\
& =f(\{0,1\},\{u_2-u_1\,u_2,2,u_2\},\{-1,\{2,m_2\},\{1,m_3\}\}) =\\ &\quad
    \norm{D}\int_{p,k}\int_{u_{1,2}}u_{2}\,\frac{p^2}
         {(k^2+m_2^2)^{2}((p+k)^2+m_3^2)}e^{-t[(-u_1
        u_2 + u_2)p^2+2k^2+u_2(p+k)^2]}
  \end{aligned}
\end{equation}
The first line contains the input for \code{ftint\_pySecDec.py}, the second
line corresponds to the notation of \cref{eq:cope}.    The
flow-time integration variables must be named \code{u1},
\code{u2}, \ldots, and the multiplication symbol ``\code{*}'' must be given
explicitly. The actual numerical values for the masses must be provided only
upon numerical integration, using the \code{---masses} option
of \code{ftint\_integrate.py}, as will be discussed in more detail
in \cref{sec:numint}.

If a propagator is massless, the mass argument is \code{0}, or it can be left
out altogether. For example, for $m_2=0$, the third argument of \verb$f$ in
the first line of \cref{eq:baor} could be given either
as \verb${-1,{2,0},{1,3}}$ or as \verb${-1,2,{1,3}}$. Recall that an integral
may contain additional infrared singularities in the limit where a mass is
zero. This is indeed the case in the example considered here: while for
$m_2\neq 0$ the integral contains only poles up to $1/\ep$
(see \cref{lst:result_0} below), there are also $1/\ep^2$ poles for
$m_2=0$. It is important to indicate all massless propagators already at the
decomposition stage in this case.

The command to run the decomposition for the integral in \cref{eq:baor},
assuming default settings, is
\begin{lstlisting}[style=shell]
  $ python3 <ftint_path>/ftint_pySecDec.py \
        'f[{0,1},{u2-u1*u2,2,u2},{-1,{2,2},{1,3}}]'
\end{lstlisting}
\code{ftint} will first convert the flow-time integral from the momentum
representation of \cref{eq:cope} to the parameter form of
\cref{eq:parameter_integral}, and then pass it on to \pysecdec.  In this
particular case, sector decomposition and subsequent compilation take of the
order of one and ten seconds on a regular modern desktop computer,
respectively.\footnote{This can change with suitable options for the compiler
and for \code{make}, see below.}  By default, the output files are written to
the directory \code{ftint\_out\_<n>}, created by \ftint\ in the current working
directory. Initially, the parameter
\code{<n>} is set to \code{0}; it is recursively increased by one if the
directory \code{ftint\_out\_<n>} already exists. In the following, we will
assume \code{<n>=0}, unless indicated otherwise.

The content of the output directory \code{ftint\_out\_0} is
\begin{lstlisting}[style=shell,caption={Contents of the output directory
      \texttt{ftint\char`_out\char`_0}.},label={lst:ftint_out}]
|-- integral_information.json
\-- secdec
    \-- secdec_ft_integral_1
        |-- secdec.out
        |-- compilation.out
        |-- ftint_data.json
        \-- disteval

\end{lstlisting}
The file \code{integral\_information.json} collects the relevant parameters of
the specific run, while the directory \code{secdec} contains one
\code{secdec\_ft\_integral\_<n>} for each compiled flow-time integral (in this case
the input consisted of only one integral). Inside there
are log-files \code{secdec.out} and \code{compilation.out} of the decomposition
and compilation respectively, a data-file \code{ftint\_data.json}
with integral-specific information, as well as the compiled integral library
\code{disteval}.

The input \code{<ft\_integrals>} in \cref{lst:deccall} can also be one or
several strings containing several flow-time integrals, or even one or several
file names. \ftint\ will extract the flow-time integrals given in the proper
format and perform the decomposition and the compilation of the integration
library for each integral (duplicates are removed).\footnote{The regular
expression searched for is
\code{f\textbackslash[\textbackslash\{*.?\textbackslash\}\textbackslash]}. The
user is advised to make sure that this cannot be confused with any other
objects in the input. See also \cref{sec:in_out}.} The output for all of them
will be written to a single output directory \code{ftint\_out\_<n>}, whose
contents will look as in \cref{lst:ftint_out}, but with a separate library
\code{secdec\_ft\_integral\_*} for each integral of the input. An example will
be presented in \cref{sec:ibp}.

\code{ftint\_pySecDec.py} provides a number of options which can be displayed
using the command
\begin{lstlisting}[style=shell]
  $ python3 <ftint_path>/ftint_pySecDec.py --help
\end{lstlisting}
These options are
\begin{description}
\item{\code{---help}:} Print this list of options.
\item{\code{---list=STR\_NUM\_LIST}:}   Specify a sublist of integrals to be
  evaluated. The format is either a comma-separated list of integers, or an
  interval of the form \code{3-10}, for example.
\item{\code{---print\_list}:} Print the list of integrals to be calculated
  and exit.
\item{\code{---exclude=EXCLUDE [EXCLUDE ...]}:} Do not evaluate the integrals
  in \code{EXCLUDE}, which is of the same format as \code{<ft\_integrals>} in
  \cref{lst:deccall}.
\item{\code{---normalform}:}  Map the integrals to their normal form.
\item{\code{---normalform\_only}:} Map the integrals to their normal form and
  stop.
\item{\code{---normalform\_list}:} Return the list of equivalent integrals and
  stop.
\item{\code{---normal\_key=NORMAL\_KEY}:}
  \code{0/-1}: define first/last element of \code{normalform\_list} as normal
  form, see \cref{sec:normalform_map}.
\item{\code{---count\_only}:} Return the number of integrals to be calculated and
  stop.
\item{\code{---latex}:} Return the input diagrams in \LaTeX\ format and stop.
\item{\code{---epsorder=EPSORDER}:} Required order in $\ep=(4-D)/2$.
\item{\code{---input\_format INPUT\_FORMAT}:}
  Specification of input format, see \cref{sec:in_out}.
\item{\code{---outdir=OUT\_DIR}:} Name of output directory.
\item{\code{---CXX\_flags=CXX\_FLAGS}:} \code{C++} compiler flags, 
  example: \code{---CXX\_flags="-mavx2 -mfma"}.
\item{\code{---make\_flags=MAKE\_FLAGS}:} \code{Makefile} flags, example: 
  \code{---make\_flags=-j4}.
\item{\code{---CUDA\_flags=CUDA\_FLAGS}:} \code{CUDA} flags to compile with
  GPU support, example: \newline \code{---CUDA\_flags=-arch=sm\_XX} where 
  \code{sm\_XX} should be replaced with the target \code{NVidia} 
  GPU architecture.
\item{\code{---integrate}:} Automatically calls \code{ftint\_integrate.py}
  after sector decomposition to perform the numerical integration (with
  default options).
\item{\code{---dimension=DIMENSION}:} The integer part $N$
  of the space-time dimension $D=N-2\,\ep$.
\item{\code{---overwrite}:} Overwrite existing directory. Otherwise, appends
  \code{\_<n>} to existing output directory.
\item{\code{---append}:} Append new integrals from the input to the compiled
  integrals in the output directory. This flag can not be passed together with
  \code{---overwrite}.
\item{\code{---ibp\_power\_goal=IBP\_POWER\_GOAL}:} Before defining subtraction 
terms, \pysecdec\ performs a number of integration by parts iterations to raise 
the order of the factorized poles to \code{IBP\_POWER\_GOAL}.
\end{description}

\subsection{Numerical integration}\label{sec:numint}

The integration part of the program is implemented in
\code{ftint\_integrate.py}. It is called from the command line as:
\begin{lstlisting}[style=shell,
  caption={Command for numerical integration.},
  label={lst:intcall}]
  $ python3 <ftint_path>/ftint_integrate.py <integral_directory> [<options>]
\end{lstlisting}
where \code{<integral\_directory>} is the name of the output directory created
by \code{ftint\_pySecDec.py}, i.e.\ \code{ftint\_out\_<n>} by default.
\ftint\ will pass the integral to \pysecdec\ for numerical integration. If not
specified otherwise, the output files will be written to the directory
\code{<integral\_directory>/result\_<m>}, where by default \code{<m>} is set to
``\code{0}'', but is recursively increased by one if the output directory
exists. The numerical result of the integral is stored in the file
\code{mathf\_out.m} in the form of a \mathematica\ replacement rule. For
example, the following command would evaluate the integral in \cref{eq:baor}
with $m_2^2=2.5/t$ and $m_3^2=3/t$ with otherwise default
settings:
\begin{lstlisting}[style=shell,
    caption={Numerical evaluation of the integral in \cref{eq:baor}.},
    label={lst:numbaor}  ]
  $ python3 <ftint_path>/ftint_integrate.py ftint_out_0 --masses=0,2.5,3
\end{lstlisting}
The numerical values for the \textit{squared} masses have to be specified via
the \code{---masses} option in units of the inverse flow time
$1/t$.\footnote{\label{foot:masses} Recall that
\cref{eq:cope} is dimensionless and thus only depends on
$m_1^2t,\cdots,m_k^2t$. It is instructive to note that, if we had specified
the l.h.s.\ of \cref{eq:baor} as
\code{f[\{0,1\},\{u2-u1*u2,2,u2\},\{-1,\{2,3\},\{1,2\}\}]}, the command in
\cref{lst:numbaor} would evaluate the r.h.s.\ of that equation with
$m_3^2=2.5/t$ and $m_2^2=3/t$.} Upon completion, \ftint\ has added a directory
\code{result\_0} to \code{ftint\_out\_0} which, aside from some other information
on the integration run, contains the file \code{mathf\_out.m} with the
following content:
\begin{lstlisting}[style=pythoncode,
  caption={Output file of the numerical integration.},label={lst:result_0}]
(*
 produced by ftint, version 1.0, Fri May 24 13:50:34 2024
*)
{
(* integral 1 [ m2**2 = 2.5/t, m3**2 = 3/t ]: *)
f[{0,1},{u2-u1*u2,2,u2},{-1,{2,2},{1,3}}] ->  (
    +eps^-2*(+0.0000000000000000*10^+00+0.0000000000000000*10^+00*I)
    +eps^-2*(+0.0000000000000000*10^+00+0.0000000000000000*10^+00*I)*plusminus
    +eps^-1*(+1.1266529421611552*10^-02+0.0000000000000000*10^+00*I)
    +eps^-1*(+2.3027953579787037*10^-10+0.0000000000000000*10^+00*I)*plusminus
    +eps^0*(+1.7797179680747620*10^-04+0.0000000000000000*10^+00*I)
    +eps^0*(+1.0762786605309655*10^-09+0.0000000000000000*10^+00*I)*plusminus
    +eps^1*(+6.8615123921316060*10^-03+0.0000000000000000*10^+00*I)
    +eps^1*(+1.1513874186746525*10^-09+0.0000000000000000*10^+00*I)*plusminus
    +eps^2*(-5.5497372202654837*10^-03+0.0000000000000000*10^+00*I)
    +eps^2*(+7.0664076906102803*10^-09+0.0000000000000000*10^+00*I)*plusminus
    +eps^3*(+5.2760242194534335*10^-03+0.0000000000000000*10^+00*I)
    +eps^3*(+3.2253823531270829*10^-08+0.0000000000000000*10^+00*I)*plusminus
  )
}
\end{lstlisting}
The integration uncertainties are marked by the variable \code{plusminus}.
The mass values are specified only in the comment on line~5. This makes it
easy to use the replacement rule in order to evaluate the same expression for
different mass values. For example, if the result of the calculation is stored
in a \mathematica\ variable \code{result} which depends on the integral under
consideration, one can obtain a numerical value for \code{result} as
\begin{lstlisting}[style=mathcode,caption={Inserting the numerical value for the
      integral within \mathematica.},label={lst:massrep}]
  replace = Get["ftint_out_0/result_0/mathf_out.m"]; result /. replace
\end{lstlisting}
One may now call \code{ftint\_integrate.py} again with different mass
parameters, and the result would be stored in
\code{ftint\_out\_0/result\_1/mathf\_out.m}. One can then use again the code in
\cref{lst:massrep}, simply replacing \code{result\_0} by \code{result\_1}.

In addition to the \mathematica\ output file, \ftint\ provides the result also
as \abbrev{YAML}\footnote{\url{https://yaml.org/}} and
\abbrev{JSON}\footnote{\url{https://www.json.org/}}
files named \code{sympyf\_out.yml} and
\code{out.json}. The \ftint\ distribution includes the files
\code{read\_yaml.py} and \code{read\_json.py} as examples on how to read these
files in \python.

To see an overview of the optional parameters related to the integration,
together with their defaults, one can run
\begin{lstlisting}[style=shell]
  $ python3 ftint_integrate.py --help
\end{lstlisting}
The options are
\begin{description}
\item{\code{---help}:} Print this list of options.
\item{\code{---masses=MASSES}:} Values for masses of each propagator, provided
  as comma-sepated list. The $n^\mathrm{th}$ value of that list is inserted
  for the mass labeled $n$ in the input of \code{ftint\_pySecDec.py}; see also
  \cref{foot:masses}.
\item{\code{---epsrel=EPSREL}:} Stop if this relative precision is
  reached.
\item{\code{---epsabs=EPSABS}:} Stop if this absolute precision is reached.
\item{\code{---delta=DELTA}:} Cut-off parameter for mass exponential,
  see \cref{eq:regulated_exponentials}.
\item{\code{---output\_format=OUTPUT\_FORMAT}:}
  Specification of output format, see \cref{sec:in_out}.
\item{\code{---points=POINTS}:} Begin integration with this lattice 
  size.\footnote{\label{QMCnote}These are QMC parameters, 
  see e.g. \citere{Heinrich:2023til} for a more detailed explanation. 
  The default settings are fine for most examples.}
\item{\code{---presamples=PRESAMPLES}:} Use this many points for 
  presampling.\footref{QMCnote}
\item{\code{---shifts=LATTICE\_SHIFTS}:} Use this many lattice shifts per
  integral.\footref{QMCnote}
\item{\code{---lattice\_candidates=LATTICE\_CANDIDATES}:} Number of median
  lattice candidates.\footref{QMCnote}
\item{\code{---outfile=OUTFILE}:} Name of the individual integration output
  files.
\item{\code{---timeout=TIMEOUT}:} The maximum number of seconds the integrator 
will spend on each integral. If \code{TIMEOUT} is reached a result that may not 
meet the desired numerical accuracy will be returned.
\end{description}
In addition, the options \code{---list}, \code{---print\_list}, \code{---overwrite} 
and \code{---outdir} are available, with the same meaning as in
\code{ftint\_pySecDec.py}, see above.

\subsection{Example: checking an integration-by-parts relation}\label{sec:ibp}

As already pointed out above, like regular Feynman integrals in dimensional
regularization, flow-time integrals obey certain \ibp\ relations which can be
derived by considering integrals over total derivatives w.r.t.\ the loop
momenta or the flow-time variables; details can be found in
\citere{Artz:2019bpr}. Let us numerically check such an \ibp\ relation in
order to give an example on how \ftint\ can be used in practice. We formulate
the relation in terms of a \mathematica\ replacement rule which we assume is
contained in a file\footnote{See \code{examples/ibp\_rule.in} in the code
repository.}
\code{ibp\_rule.in}, see \cref{lst:ibprule}.
\begin{lstlisting}[style=mathcode,caption={A \three-loop \ibp\ identity in
      \texttt{Mathematica} format.},label={lst:ibprule}]
{f[{},{0,0,0,1,1,1},{-1,1,1,1,1,1}] -> 
-f[{},{0,0,0,1,1,1},{0,1,1,1,1,0}]/2
-(f[{},{0,0,0,1,1,1},{1,0,1,1,0,0}])/(1-n/4)
-(f[{},{0,0,0,1,1,1},{1,1,0,0,0,0}])/(2*(1+(-7/12+n/12)*n))}
\end{lstlisting}
Here, \code{n}$=4-2\,\ep$.  Since some of the integrals on the r.h.s.\ have a
prefactor $\sim 1/\ep$, we need to evaluate them through \order{\ep} in order
to check this relation through \order{\ep^0}, while the other integrals are
needed only through \order{\ep^0}. In realistic cases, it is advisable to
split the set of integrals according to the required power in $\ep$. For this
simple example though, we evaluate all integrals to \order{\ep}.

This is done by first performing the sector decomposition:
\begin{lstlisting}[style=shell]
  $ python3 <ftint_path>/ftint_pySecDec.py ibp_rule.in --epsorder=1
\end{lstlisting}
This will first report that \ftint\ finds four different flow-time integrals
in the file. It will then perform the sector decomposition for the first
diagram and compile the corresponding integration library, before turning to
the second diagram etc. After completion, \ftint\ has created a directory
named \code{ftint\_out\_0} with the following structure:
\begin{lstlisting}[style=shell]
\-- ftint_out_0
    |-- integral_information.json
    \-- secdec
        |-- secdec_ft_integral_1
        |-- secdec_ft_integral_2
        |-- secdec_ft_integral_3
        \-- secdec_ft_integral_4
\end{lstlisting}
The directories \code{secdec\_ft\_integral\_*} contain the integration libraries
for each of the four integrals, as well as other information required by
\ftint\ for the numerical integration.  The latter is performed through the
command
\begin{lstlisting}[style=shell]
  $ python3 <ftint_path>/ftint_integrate.py ftint_out_0
\end{lstlisting}
This again first reports that four integrals will be evaluated. The
corresponding numerical results will be printed to the screen. After
completion, \ftint\ will have created a subdirectory named \code{results\_0} in
\code{ftint\_out\_0}, which, among other information, contains the file
\code{mathf\_out.m}, whose contents are shown in \cref{lst:ibpresult}.
\begin{lstlisting}[style=pythoncode,caption={Result for the integrals of
      \cref{lst:ibprule}. Only the first few lines are shown.},
  label={lst:ibpresult}]
(*
 produced by ftint, version 1.0, Fri May 24 09:21:05 2024
*)
{
(* integral 1 : *)
f[{},{0,0,0,1,1,1},{-1,1,1,1,1,1}] ->  (
    +eps^-1*(+1.4384102482242656*10^-01+0.0000000000000000*10^+00*I)
    +eps^-1*(+7.5901486511209429*10^-09+0.0000000000000000*10^+00*I)*plusminus
    +eps^0*(+9.1030039859574119*10^-01+0.0000000000000000*10^+00*I)
    +eps^0*(+7.7645751796859680*10^-06+0.0000000000000000*10^+00*I)*plusminus
    +eps^1*(+3.6458111371273918*10^+00+0.0000000000000000*10^+00*I)
    +eps^1*(+4.2456809764811359*10^-05+0.0000000000000000*10^+00*I)*plusminus
  ),
(* integral 2 : *)
f[{},{0,0,0,1,1,1},{0,1,1,1,1,0}] ->  (
    +eps^-1*(+2.8768207244049038*10^-01+0.0000000000000000*10^+00*I)
    +eps^-1*(+1.2597382329360634*10^-11+0.0000000000000000*10^+00*I)*plusminus
    +eps^0*(+1.4717502294126590*10^+00+0.0000000000000000*10^+00*I)
    .
    .
    .
\end{lstlisting}
One can now check the \ibp\ relation within \mathematica\ using the following
code:\footnote{See \code{examples/check\_ibp\_rule.m} in the code repository.}
\begin{lstlisting}[style=mathcode]
rule = Get["ibp_rule.in"][[1]];
replace = Get["ftint_out_0/result_0/mathf_out.m"];
check = Normal[Series[(rule[[1]]-rule[[2]]) /. n -> 4-2*eps /. replace,
		      {eps,0,0}]];
\end{lstlisting}
The result is
\begin{lstlisting}[style=mathcode]
  
                    -15             -15
         -3.66374 10    + 4.85345 10    plusminus
Out[19]= ---------------------------------------- +
                              2
                           eps

               -9
     7.64685 10   (-1.48995 + 1. plusminus)
>    -------------------------------------- +
                      eps

               -6
>    7.90387 10   (2.47724 + 1. plusminus)
\end{lstlisting}
meaning that the l.h.s.\ and the r.h.s.\ of the relation agree within the
default numerical precision. It is now easy to increase this precision by
running
\begin{lstlisting}[style=shell]
  $ python3 <ftint_path>/ftint_integrate.py ftint_out_0 \
        --epsrel=1e-8 --epsabs=1e-8
\end{lstlisting}

\subsection{Mapping to the normal form}\label{sec:normalform_map}

If one needs to compute a large list of integrals, it may be advantageous to
map them to their normal form before integration,
see \cref{sec:normalform}. This can be done by
calling \code{ftint\_pySecDec.py} with the option \code{---normalform}. \ftint\
will then produce the file \code{ftint\_out\_<n>/normalmap.m} which contains the
mapping of each integral to its normal form (unless it already is in the
normal form) in \mathematica\ format. \ftint\ will then proceed with the
calculation only for the normal-form integrals.

For example, if the file\footnote{See \code{examples/normal\_form.in} in the code repository.} \code{normal\_form.in} contains the following list of
flow-time integrals
\begin{lstlisting}[style=mathcode,caption={List of flow-time integrals to be
brought to normal form.},label={lst:normal}]
f[{},{2,1,0},{2,2,1}]
f[{1},{2*u1,u1,u1},{2,1,3}]
f[{2,1},{u1*u2,2,u1},{3,2,1}]
f[{1,2},{u1*u2,u2,2},{3,1,2}]
\end{lstlisting}
then the call
\begin{lstlisting}[style=shell]
  $ python3 <ftint_path>/ftint_pySecDec.py integrals.m --normalform
\end{lstlisting}
will produce the file \code{ftint\_out\_0/normal\_form.in} with the following
content:
\begin{lstlisting}[style=mathcode]
{f[{},{2,1,0},{2,2,1}] -> f[{},{2,1,0},{{2,0},{2,0},{1,0}}],
f[{1},{2*u1,u1,u1},{2,1,3}] -> f[{1},{2*u1,u1,u1},{{2,0},{3,0},{1,0}}],
f[{2,1},{u1*u2,2,u1},{3,2,1}] -> f[{2,1},{u1*u2,u1,2},{{3,0},{1,0},{2,0}}],
f[{1,2},{u1*u2,u2,2},{3,1,2}] -> f[{2,1},{u1*u2,u1,2},{{3,0},{1,0},{2,0}}]}
\end{lstlisting}
The first thing to notice is that \ftint\ does not use the abbreviated
notation for massless propagators, simply to ensure a unique output format.
Furthermore, aside from this notational aspect, the integral in line~1 of
\cref{lst:normal} is already in normal form. The integrals in line~3 and
line~4 are mapped to the same normal form. Thus, \ftint\ will do the sector
decomposition only for the \textit{three} different normal-form integrals.  As
usual, it will write the integration libraries to \code{ftint\_out\_0}, compile
them, and the user can evaluate them numerically using
\code{ftint\_integrate.py}.

If the user is only interested in the normal-form mappings, one may use
the \code{---normalform\_only} option instead. In this case, \ftint\ will
write the file \code{ftint\_out\_0/normalform.m} and stop.

For integrals with non-vanishing masses, the procedure works in the very same
way. The only subtlety here is that the order of the masses may
change. Consider, for example, the integral
\begin{equation}\label{eq:ibp:engr}
  \begin{aligned}
\verb$f[{1,2},{3*u1,2,u1*u2},{{1,1},{4,2},{2,3}}]$
  =\\ =\int_{p,k}\int_{u_{1,2}}u_{1}u_{2}^{2}\,\frac{e^{-t[3 u_1p^2+2k^2+u_1
        u_2(p+k)^2]}}{(p^2+m_1^2)(k^2+m_2^2)^{4}((p+k)^2+m_3^2)^{2}}\,.
  \end{aligned}
\end{equation}
Mapping it to normal form will result in
\begin{equation}\label{eq:ibp:kama}
  \begin{aligned}
    \verb$f[{1,2},{u1*u2,3*u1,2},{{2,3},{1,1},{4,2}}]$ =\\ \quad
    =\int_{p,k}\int_{u_{1,2}}u_{1}u_{2}^{2}\,\frac{e^{-t[u_1 u_2p^2+3
          u_1k^2+2(p+k)^2]}}{(p^2+m_3^2)^{2}(k^2+m_1^2)((p+k)^2+m_2^2)^{4}}\,.
  \end{aligned}
\end{equation}
In order to numerically evaluate these integrals with
\code{ftint\_integrate.py}, one must use the same order of arguments
$m_1,m_2,m_3$ in the \code{---masses} option. This means that the command for
the integration is independent of whether one evaluates the original integral
or its normal form (aside from the fact that their integration libraries may be
located in different directories). 

As described in \cref{sec:normalform}, if the option \code{---normalform} is
given, \ftint\ generates a list of equivalent integrals and continues the
calculation with the \textit{first} element of this list. The user may alter
this behavior by adding the option \code{---normal\_key=-1}, in which
case \ftint\ will continue with the \textit{last} element of the sorted
list. The full (sorted) list can be viewed by calling \ftint\ with the
option \code{---normalform\_list}.  We have not observed any significant
differences in computing times among these integrals.

\subsection{Checks}

\subsubsection{Analytic solutions, symmetries, and simple identities}

There are a number of rather straightforward checks that we have used to
validate \ftint:
\begin{itemize}
\item
  Some simple integrals can be solved analytically, see, e.g.,
  \cref{eq:hire,eq:massive:gens}. We have compared a number of them to the
  numerical result from \ftint\ and found agreement.
\item An integral and its normal form, see \cref{sec:normalform}, must lead to
the same numerical result, of course. We have confirmed this symmetry
with \ftint, which both checks the numerical evaluation of the integral, as
well as the algorithm and implementation for mapping the integrals to their
normal form. 
\item 
Certain multi-loop integrals can be written as products of integrals at lower
loop order. For example, it is easy to see that
\begin{equation}\label{eq:food}
  \begin{aligned}
    f(\{\},\{1,0,1\},\{-1,0,0\}) = f(\{\},\{1\},\{-1\})\cdot f(\{\},\{1\},\{0\})
  \end{aligned}
\end{equation}
All such checks were passed by \ftint.
\end{itemize}

\subsubsection{Flow-time derivatives}

In a dimensionally regularized integral, one can interchange integration and
differentiation without changing the result. In the massless case, one may
derive non-trivial relations from this that can be easily checked. For
example, since the integral as defined in \cref{eq:cope} is dimensionless, in
the massless case it follows that
\begin{equation}\label{eq:idyl}
  \begin{aligned}
    0 = t\deriv{}{}{t} f(\mathbf{c},\mathbf{a},\mathbf{b})
    &= \left(\frac{lD}{2}-b\right) f(\mathbf{c},\mathbf{a},\mathbf{b})\\&
    - \norm{lD/2}\,t^{-b+1}\int_{[0,1]^f} \dd
    \mathbf{u}\,\mathbf{u}^\mathbf{c}\sum_{j=1}^k a_j \int_{\mathbf{p}} P_j^2\,
    \frac{\exp[-t\,\sum_{i=1}^k a_iP_i^2]}
         {(P_1^2)^{b_1}\cdots(P_k^2)^{b_k}}\,.
  \end{aligned}
\end{equation}
Cancelling the power of $P_j^2$ and inserting the explicit polynomials $a_j$
of the flow-time variables turns the r.h.s.\ into the sum of regular flow-time
integrals of the form \cref{eq:cope} with varied parameters. As an example,
consider again a two-loop case:
\begin{equation}\label{eq:hare}
  \begin{aligned}
    (-6+2\,\ep)\,&f(\{\}, \{1, 1, 0\}, \{0, -2, 0\})=\\&= -f(\{\}, \{1, 1,
    0\}, \{-1, -2, 0\}) - f(\{\}, \{1, 1, 0\}, \{0, -3, 0\})\,.
  \end{aligned}
\end{equation}
This relation can be checked in close analogy to the example discussed in
\cref{sec:ibp}.

\subsubsection{Integration-by-parts identities}

In \cref{sec:ibp}, we used the check of a two-loop \ibp\ relation in order to
demonstrate the operation of \ftint. In fact, we have used hundreds of such
relations at three-loop level, derived in the context of calculations
performed in \citere{Harlander:2020duo}, for example, in order to
check \ftint.

\section{Conclusions and Outlook}\label{sec:conclusions}

With more and more potential applications of the perturbative approach to the
\gff\ identified, the demand for suitable software tools is increasing. In
this paper, we described the application of the sector-decomposition algorithm
to flow-time integrals up to the \three-loop level in the form of a
\python\ program named \ftint. It transforms a flow-time integral without
external momenta to a multidimensional parameter integral over a unit
hypercube, which is then passed to the public library \pysecdec\ for sector
decomposition and numerical integration.

We have performed a number of checks on the program and made an effort towards
user-friendliness and flexibility. Future releases of the program will support
partial fractioning for propagators with identical momenta but different
masses, as well as non-vanishing external momenta. Furthermore, we plan to
make use of \pySecDec's main strengths of evaluating complete amplitudes
rather than individual integrals.

\section*{Acknowledgments}
We would like to thank Janosch Borgulat, Nils Felten, Gudrun Heinrich,
Stephen Jones, Matthias Kerner, Jonas Kohnen, Fabian Lange, Henri Lindlahr, 
Vitaly Magerya, Tobias Neumann, Johannes Schlenk, and Henry Werthenbach for 
helpful input and comments.  This research was supported by the \textit{Deutsche
Forschungsgemeinschaft (DFG, German Research Foundation) under grants
396021762 - TRR 257} and \textit{460791904}.

\begin{appendix}
  \section{In- and output format}\label{sec:in_out}
  
  By default, \ftint\ assumes the format defined in \cref{eq:cope}, as
  exemplified by \cref{eq:baor}, for the flow-time integrals, both in the
  input and the output. \ftint\ provides a basic way to convert between other
  formats and the \ftint-format by editing the file
  \code{user\_format.py}. This defines three functions. For the sake of
  clarity, let us assume that the user would like to perform a mapping
\begin{lstlisting}[style=mathcode]
f[{a},{b},{c}] <-> g([a],[b],[c])
\end{lstlisting}
  The purpose of the three functions in \code{user\_format.py} is given as
  follows.
  \begin{itemize}
  \item\code{user\_patterns} allows the user to specify the pattern which
    defines a gradient-flow integral. In the example above, the function could
    be defined as
\begin{lstlisting}[style=pythoncode]
def user_patterns(input_format):
    if input_format==2:
        out = [r"g\(\[.*?\]\)"]
    else:
        out = [r"f\[\{.*?\}\]"]
    return(out)
\end{lstlisting}
  \item\code{from\_user} defines how to translate the user's format to the
    \ftint\ format. In this case, one could define
\begin{lstlisting}[style=pythoncode]
def from_user(input_format,string):
    if input_format==2:
        out = re.sub(r'g\(\[(.*?)\],\[(.+?)\],\[(.+?)\]\)',r'f[{\1},{\2},{\3}]',string)
    else:
        out = string
    return(out)
\end{lstlisting}
  \item\code{to\_user} defines how to translate the \ftint\ format to the
    user's format. For the current example, this could be achieved through
\begin{lstlisting}[style=pythoncode]
def to_user(output_format,string):
    if output_format==2:
        out = re.sub(r'f\[\{(.*?)\},\{(.+?)\},\{(.+?)\}\]', r'g([\1],[\2],[\3])',string)
    else:
        out = string
    return(out)
\end{lstlisting}
  \end{itemize}
  The user can now switch to the new input format by providing the option
  \code{---input\_format=2} to \code{ftint\_pySecDec.py}. The new output
  format is obtained by providing the option \code{---output\_format=2} to
  \code{ftint\_integrate.py}. Without these options, the default format will
  be adopted.  Let us stress that this is a very rudimentary
  implementation. The user is advised to use it with care. It may be safer to
  convert all input to the default \ftint\ notation before using \ftint.
\end{appendix}

\bibliographystyle{utphys}
\bibliography{paper}

\providecommand{\href}[2]{#2}\begingroup\raggedright\begin{thebibliography}{10}

\bibitem{Narayanan:2006rf}
R.~Narayanan and H.~Neuberger, {\emph{Infinite N phase transitions in continuum
  Wilson loop operators}},
  \href{http://dx.doi.org/10.1088/1126-6708/2006/03/064}{{\em JHEP} {\bfseries
  03} (2006) 064}, \href{http://arxiv.org/abs/hep-th/0601210}{{\ttfamily
  arXiv:hep-th/0601210}}.

\bibitem{Luscher:2009eq}
M.~L\"uscher, {\emph{Trivializing maps, the Wilson flow and the HMC
  algorithm}}, \href{http://dx.doi.org/10.1007/s00220-009-0953-7}{{\em Commun.
  Math. Phys.} {\bfseries 293} (2010) 899--919},
  \href{http://arxiv.org/abs/0907.5491}{{\ttfamily arXiv:0907.5491 [hep-lat]}}.

\bibitem{Luscher:2010iy}
M.~L\"uscher, {\emph{Properties and uses of the Wilson flow in lattice QCD}},
  \href{http://dx.doi.org/10.1007/JHEP08(2010)071}{{\em JHEP} {\bfseries 08}
  (2010) 071}, \href{http://arxiv.org/abs/1006.4518}{{\ttfamily arXiv:1006.4518
  [hep-lat]}}. [Erratum: JHEP 03, 092 (2014)].

\bibitem{Luscher:2011bx}
M.~L\"uscher and P.~Weisz, {\emph{Perturbative analysis of the gradient flow in
  non-abelian gauge theories}},
  \href{http://dx.doi.org/10.1007/JHEP02(2011)051}{{\em JHEP} {\bfseries 02}
  (2011) 051}, \href{http://arxiv.org/abs/1101.0963}{{\ttfamily arXiv:1101.0963
  [hep-th]}}.

\bibitem{Luscher:2013cpa}
M.~L\"uscher, {\emph{Chiral symmetry and the Yang--Mills gradient flow}},
  \href{http://dx.doi.org/10.1007/JHEP04(2013)123}{{\em JHEP} {\bfseries 04}
  (2013) 123}, \href{http://arxiv.org/abs/1302.5246}{{\ttfamily arXiv:1302.5246
  [hep-lat]}}.

\bibitem{BMW:2012hcm}
BMW collaboration, S.~Bors\'anyi, S.~D\"urr, Z.~Fodor, C.~Hoelbling, S.~D.
  Katz, S.~Krieg, T.~Kurth, L.~Lellouch, T.~Lippert, and C.~McNeile,
  {\emph{High-precision scale setting in lattice QCD}},
  \href{http://dx.doi.org/10.1007/JHEP09(2012)010}{{\em JHEP} {\bfseries 09}
  (2012) 010}, \href{http://arxiv.org/abs/1203.4469}{{\ttfamily arXiv:1203.4469
  [hep-lat]}}.

\bibitem{Suzuki:2013gza}
H.~Suzuki, {\emph{Energy\textendash{}momentum tensor from the
  Yang\textendash{}Mills gradient flow}},
  \href{http://dx.doi.org/10.1093/ptep/ptt059}{{\em PTEP} {\bfseries 2013}
  (2013) 083B03}, \href{http://arxiv.org/abs/1304.0533}{{\ttfamily
  arXiv:1304.0533 [hep-lat]}}. [Erratum: PTEP 2015, 079201 (2015)].

\bibitem{Makino:2014taa}
H.~Makino and H.~Suzuki, {\emph{Lattice energy\textendash{}momentum tensor from
  the Yang\textendash{}Mills gradient flow\textemdash{}inclusion of fermion
  fields}}, \href{http://dx.doi.org/10.1093/ptep/ptu070}{{\em PTEP} {\bfseries
  2014} (2014) 063B02}, \href{http://arxiv.org/abs/1403.4772}{{\ttfamily
  arXiv:1403.4772 [hep-lat]}}. [Erratum: PTEP 2015, 079202 (2015)].

\bibitem{Iritani:2018idk}
T.~Iritani, M.~Kitazawa, H.~Suzuki, and H.~Takaura, {\emph{Thermodynamics in
  quenched QCD: energy\textendash{}momentum tensor with two-loop order
  coefficients in the gradient-flow formalism}},
  \href{http://dx.doi.org/10.1093/ptep/ptz001}{{\em PTEP} {\bfseries 2019}
  no.~2, (2019) 023B02}, \href{http://arxiv.org/abs/1812.06444}{{\ttfamily
  arXiv:1812.06444 [hep-lat]}}.

\bibitem{Harlander:2018zpi}
R.~V. Harlander, Y.~Kluth, and F.~Lange, {\emph{The two-loop
  energy\textendash{}momentum tensor within the gradient-flow formalism}},
  \href{http://dx.doi.org/10.1140/epjc/s10052-018-6415-7}{{\em Eur. Phys. J. C}
  {\bfseries 78} no.~11, (2018) 944},
  \href{http://arxiv.org/abs/1808.09837}{{\ttfamily arXiv:1808.09837
  [hep-lat]}}. [Erratum: Eur.Phys.J.C 79, 858 (2019)].

\bibitem{Suzuki:2020zue}
A.~Suzuki, Y.~Taniguchi, H.~Suzuki, and K.~Kanaya, {\emph{Four quark operators
  for kaon bag parameter with gradient flow}},
  \href{http://dx.doi.org/10.1103/PhysRevD.102.034508}{{\em Phys. Rev. D}
  {\bfseries 102} no.~3, (2020) 034508},
  \href{http://arxiv.org/abs/2006.06999}{{\ttfamily arXiv:2006.06999
  [hep-lat]}}.

\bibitem{Suzuki:2021tlr}
H.~Suzuki and H.~Takaura, {\emph{$t \to 0$ extrapolation function in the small
  flow time expansion method for the energy\textendash{}momentum tensor}},
  \href{http://dx.doi.org/10.1093/ptep/ptab068}{{\em PTEP} {\bfseries 2021}
  no.~7, (2021) 073B02}, \href{http://arxiv.org/abs/2102.02174}{{\ttfamily
  arXiv:2102.02174 [hep-lat]}}.

\bibitem{Harlander:2022tgk}
R.~V. Harlander and F.~Lange, {\emph{Effective electroweak Hamiltonian in the
  gradient-flow formalism}},
  \href{http://dx.doi.org/10.1103/PhysRevD.105.L071504}{{\em Phys. Rev. D}
  {\bfseries 105} no.~7, (2022) L071504},
  \href{http://arxiv.org/abs/2201.08618}{{\ttfamily arXiv:2201.08618
  [hep-lat]}}.

\bibitem{Black:2023vju}
M.~Black, R.~Harlander, F.~Lange, A.~Rago, A.~Shindler, and O.~Witzel,
  {\emph{Using gradient flow to renormalise matrix elements for meson mixing
  and lifetimes}}, \href{http://dx.doi.org/10.22323/1.453.0263}{{\em PoS}
  {\bfseries LATTICE2023} (2024) 263},
  \href{http://arxiv.org/abs/2310.18059}{{\ttfamily arXiv:2310.18059
  [hep-lat]}}.

\bibitem{Rizik:2020naq}
SymLat collaboration, M.~D. Rizik, C.~J. Monahan, and A.~Shindler, {\emph{Short
  flow-time coefficients of $CP$-violating operators}},
  \href{http://dx.doi.org/10.1103/PhysRevD.102.034509}{{\em Phys. Rev. D}
  {\bfseries 102} no.~3, (2020) 034509},
  \href{http://arxiv.org/abs/2005.04199}{{\ttfamily arXiv:2005.04199
  [hep-lat]}}.

\bibitem{Harlander:2020duo}
R.~V. Harlander, F.~Lange, and T.~Neumann, {\emph{Hadronic vacuum polarization
  using gradient flow}}, \href{http://dx.doi.org/10.1007/JHEP08(2020)109}{{\em
  JHEP} {\bfseries 08} (2020) 109},
  \href{http://arxiv.org/abs/2007.01057}{{\ttfamily arXiv:2007.01057
  [hep-lat]}}.

\bibitem{Mereghetti:2021nkt}
E.~Mereghetti, C.~J. Monahan, M.~D. Rizik, A.~Shindler, and P.~Stoffer,
  {\emph{One-loop matching for quark dipole operators in a gradient-flow
  scheme}}, \href{http://dx.doi.org/10.1007/JHEP04(2022)050}{{\em JHEP}
  {\bfseries 04} (2022) 050}, \href{http://arxiv.org/abs/2111.11449}{{\ttfamily
  arXiv:2111.11449 [hep-lat]}}.

\bibitem{Harlander:2022vgf}
R.~Harlander, M.~D. Rizik, J.~Borgulat, and A.~Shindler, {\emph{Two-loop
  matching of the chromo-magnetic dipole operator with the gradient flow}},
  \href{http://dx.doi.org/10.22323/1.430.0313}{{\em PoS} {\bfseries
  LATTICE2022} (2023) 313}, \href{http://arxiv.org/abs/2212.09824}{{\ttfamily
  arXiv:2212.09824 [hep-lat]}}.

\bibitem{Borgulat:2023xml}
J.~Borgulat, R.~V. Harlander, J.~T. Kohnen, and F.~Lange,
  {\emph{Short-flow-time expansion of quark bilinears through
  next-to-next-to-leading order QCD}},
  \href{http://dx.doi.org/10.1007/JHEP05(2024)179}{{\em JHEP} {\bfseries 05}
  (2024) 179}, \href{http://arxiv.org/abs/2311.16799}{{\ttfamily
  arXiv:2311.16799 [hep-lat]}}.

\bibitem{Shindler:2023xpd}
A.~Shindler, {\emph{Moments of parton distribution functions of any order from
  lattice QCD}}, \href{http://arxiv.org/abs/2311.18704}{{\ttfamily
  arXiv:2311.18704 [hep-lat]}}.

\bibitem{Dragos:2019oxn}
J.~Dragos, T.~Luu, A.~Shindler, J.~de~Vries, and A.~Yousif, {\emph{Confirming
  the existence of the strong CP problem in lattice QCD with the gradient
  flow}}, \href{http://dx.doi.org/10.1103/PhysRevC.103.015202}{{\em Phys. Rev.
  C} {\bfseries 103} no.~1, (2021) 015202},
  \href{http://arxiv.org/abs/1902.03254}{{\ttfamily arXiv:1902.03254
  [hep-lat]}}.

\bibitem{Artz:2019bpr}
J.~Artz, R.~V. Harlander, F.~Lange, T.~Neumann, and M.~Prausa, {\emph{Results
  and techniques for higher order calculations within the gradient-flow
  formalism}}, \href{http://dx.doi.org/10.1007/JHEP06(2019)121}{{\em JHEP}
  {\bfseries 06} (2019) 121}, \href{http://arxiv.org/abs/1905.00882}{{\ttfamily
  arXiv:1905.00882 [hep-lat]}}. [Erratum: JHEP 10, 032 (2019)].

\bibitem{Borowka:2017idc}
S.~Borowka, G.~Heinrich, S.~Jahn, S.~P. Jones, M.~Kerner, J.~Schlenk, and
  T.~Zirke, {\emph{pySecDec: a toolbox for the numerical evaluation of
  multi-scale integrals}},
  \href{http://dx.doi.org/10.1016/j.cpc.2017.09.015}{{\em Comput. Phys.
  Commun.} {\bfseries 222} (2018) 313--326},
  \href{http://arxiv.org/abs/1703.09692}{{\ttfamily arXiv:1703.09692
  [hep-ph]}}.

\bibitem{Borowka:2018goh}
S.~Borowka, G.~Heinrich, S.~Jahn, S.~P. Jones, M.~Kerner, and J.~Schlenk,
  {\emph{A GPU compatible quasi-Monte Carlo integrator interfaced to
  pySecDec}}, \href{http://dx.doi.org/10.1016/j.cpc.2019.02.015}{{\em Comput.
  Phys. Commun.} {\bfseries 240} (2019) 120--137},
  \href{http://arxiv.org/abs/1811.11720}{{\ttfamily arXiv:1811.11720
  [physics.comp-ph]}}.

\bibitem{Heinrich:2023til}
G.~Heinrich, S.~P. Jones, M.~Kerner, V.~Magerya, A.~Olsson, and J.~Schlenk,
  {\emph{Numerical scattering amplitudes with pySecDec}},
  \href{http://dx.doi.org/10.1016/j.cpc.2023.108956}{{\em Comput. Phys.
  Commun.} {\bfseries 295} (2024) 108956},
  \href{http://arxiv.org/abs/2305.19768}{{\ttfamily arXiv:2305.19768
  [hep-ph]}}.

\bibitem{Binoth:2000ps}
T.~Binoth and G.~Heinrich, {\emph{An automatized algorithm to compute infrared
  divergent multiloop integrals}},
  \href{http://dx.doi.org/10.1016/S0550-3213(00)00429-6}{{\em Nucl. Phys. B}
  {\bfseries 585} (2000) 741--759},
  \href{http://arxiv.org/abs/hep-ph/0004013}{{\ttfamily arXiv:hep-ph/0004013}}.

\bibitem{Binoth:2003ak}
T.~Binoth and G.~Heinrich, {\emph{Numerical evaluation of multiloop integrals
  by sector decomposition}},
  \href{http://dx.doi.org/10.1016/j.nuclphysb.2003.12.023}{{\em Nucl. Phys. B}
  {\bfseries 680} (2004) 375--388},
  \href{http://arxiv.org/abs/hep-ph/0305234}{{\ttfamily arXiv:hep-ph/0305234}}.

\bibitem{Heinrich:2008si}
G.~Heinrich, {\emph{Sector decomposition}},
  \href{http://dx.doi.org/10.1142/S0217751X08040263}{{\em Int. J. Mod. Phys. A}
  {\bfseries 23} (2008) 1457--1486},
  \href{http://arxiv.org/abs/0803.4177}{{\ttfamily arXiv:0803.4177 [hep-ph]}}.

\bibitem{Harlander:2016vzb}
R.~V. Harlander and T.~Neumann, {\emph{The perturbative QCD gradient flow to
  three loops}}, \href{http://dx.doi.org/10.1007/JHEP06(2016)161}{{\em JHEP}
  {\bfseries 06} (2016) 161}, \href{http://arxiv.org/abs/1606.03756}{{\ttfamily
  arXiv:1606.03756 [hep-ph]}}.

\bibitem{Gorishnii:1983su}
S.~G. Gorishnii, S.~A. Larin, and F.~V. Tkachov, {\emph{The algorithm for OPE
  coefficient functions in the MS scheme}},
  \href{http://dx.doi.org/10.1016/0370-2693(83)91439-9}{{\em Phys. Lett. B}
  {\bfseries 124} (1983) 217--220}.

\bibitem{Gorishnii:1986gn}
S.~G. Gorishnii and S.~A. Larin, {\emph{Coefficient functions of asymptotic
  operator expansions in minimal subtraction scheme}},
  \href{http://dx.doi.org/10.1016/0550-3213(87)90283-5}{{\em Nucl. Phys. B}
  {\bfseries 283} (1987) 452}.

\bibitem{Harlander:2020cyh}
R.~V. Harlander, S.~Y. Klein, and M.~Lipp, {\emph{FeynGame}},
  \href{http://dx.doi.org/10.1016/j.cpc.2020.107465}{{\em Comput. Phys.
  Commun.} {\bfseries 256} (2020) 107465},
  \href{http://arxiv.org/abs/2003.00896}{{\ttfamily arXiv:2003.00896
  [physics.ed-ph]}}.

\bibitem{Harlander:2024qbn}
R.~Harlander, S.~Y. Klein, and M.~C. Schaaf, {\emph{FeynGame-2.1 -- Feynman
  diagrams made easy}}, \href{http://dx.doi.org/10.22323/1.449.0657}{{\em PoS}
  {\bfseries EPS-HEP2023} (2024) 657},
  \href{http://arxiv.org/abs/2401.12778}{{\ttfamily arXiv:2401.12778
  [hep-ph]}}.

\bibitem{Smirnov:2008py}
A.~V. Smirnov and M.~N. Tentyukov, {\emph{Feynman Integral Evaluation by a
  Sector decomposiTion Approach (FIESTA)}},
  \href{http://dx.doi.org/10.1016/j.cpc.2008.11.006}{{\em Comput. Phys.
  Commun.} {\bfseries 180} (2009) 735--746},
  \href{http://arxiv.org/abs/0807.4129}{{\ttfamily arXiv:0807.4129 [hep-ph]}}.

\bibitem{Smirnov:2009pb}
A.~V. Smirnov, V.~A. Smirnov, and M.~Tentyukov, {\emph{FIESTA 2:
  Parallelizeable multiloop numerical calculations}},
  \href{http://dx.doi.org/10.1016/j.cpc.2010.11.025}{{\em Comput. Phys.
  Commun.} {\bfseries 182} (2011) 790--803},
  \href{http://arxiv.org/abs/0912.0158}{{\ttfamily arXiv:0912.0158 [hep-ph]}}.

\bibitem{Smirnov:2013eza}
A.~V. Smirnov, {\emph{FIESTA 3: cluster-parallelizable multiloop numerical
  calculations in physical regions}},
  \href{http://dx.doi.org/10.1016/j.cpc.2014.03.015}{{\em Comput. Phys.
  Commun.} {\bfseries 185} (2014) 2090--2100},
  \href{http://arxiv.org/abs/1312.3186}{{\ttfamily arXiv:1312.3186 [hep-ph]}}.

\bibitem{Smirnov:2015mct}
A.~V. Smirnov, {\emph{FIESTA4: Optimized Feynman integral calculations with GPU
  support}}, \href{http://dx.doi.org/10.1016/j.cpc.2016.03.013}{{\em Comput.
  Phys. Commun.} {\bfseries 204} (2016) 189--199},
  \href{http://arxiv.org/abs/1511.03614}{{\ttfamily arXiv:1511.03614
  [hep-ph]}}.

\bibitem{Kaneko:2009qx}
T.~Kaneko and T.~Ueda, {\emph{A geometric method of sector decomposition}},
  \href{http://dx.doi.org/10.1016/j.cpc.2010.04.001}{{\em Comput. Phys.
  Commun.} {\bfseries 181} (2010) 1352--1361},
  \href{http://arxiv.org/abs/0908.2897}{{\ttfamily arXiv:0908.2897 [hep-ph]}}.

\bibitem{Heinrich:2021dbf}
G.~Heinrich, S.~Jahn, S.~P. Jones, M.~Kerner, F.~Langer, V.~Magerya,
  A.~P\"oldaru, J.~Schlenk, and E.~Villa, {\emph{Expansion by regions with
  pySecDec}}, \href{http://dx.doi.org/10.1016/j.cpc.2021.108267}{{\em Comput.
  Phys. Commun.} {\bfseries 273} (2022) 108267},
  \href{http://arxiv.org/abs/2108.10807}{{\ttfamily arXiv:2108.10807
  [hep-ph]}}.

\bibitem{Mathematica}
{Wolfram Research, Inc.}, ``Mathematica, {V}ersion 14.1.''
\newblock \url{https://www.wolfram.com/mathematica}. Champaign, IL, 2024.

\bibitem{Vermaseren:2000nd}
J.~A.~M. Vermaseren, {\emph{New features of FORM}},
  \href{http://arxiv.org/abs/math-ph/0010025}{{\ttfamily
  arXiv:math-ph/0010025}}.

\bibitem{Kuipers:2012rf}
J.~Kuipers, T.~Ueda, J.~A.~M. Vermaseren, and J.~Vollinga, {\emph{FORM version
  4.0}}, \href{http://dx.doi.org/10.1016/j.cpc.2012.12.028}{{\em Comput. Phys.
  Commun.} {\bfseries 184} (2013) 1453--1467},
  \href{http://arxiv.org/abs/1203.6543}{{\ttfamily arXiv:1203.6543 [cs.SC]}}.

\end{thebibliography}\endgroup

\end{document}